\documentclass[fleqn,usenatbib]{mnras}

\usepackage{newtxtext,newtxmath}

\usepackage[T1]{fontenc}

\DeclareRobustCommand{\VAN}[3]{#2}
\let\VANthebibliography\thebibliography
\def\thebibliography{\DeclareRobustCommand{\VAN}[3]{##3}\VANthebibliography}

\usepackage{graphicx}	
\usepackage{amsmath}	

\title[Pyssed]{PySSED: an automated method of collating and fitting stellar spectral energy distributions}

\author[I. McDonald et al.]{
Iain McDonald,$^{1}$\thanks{E-mail: iain.mcdonald-2@manchester.ac.uk}
Albert A. Zijlstra,$^{1}$
Nick L. J. Cox,$^{2}$ Emma L. Alexander,$^{1}$ 
Alexander Csukai,$^{1}$ \newauthor Ria Ramkumar$^{1}$ and Alexander Hollings$^{1}$
\\
$^{1}$Jodrell Bank Centre for Astrophysics, Department of Physics and Astronomy, The University of Manchester, Oxford Road, M13 9PL, UK\\
$^{2}$ACRI-ST, Centre d'Etudes et de Recherche de Grasse (CERGA), 10 Av. Nicolas Copernic, 06130, Grasse, France
}

\date{Accepted XXX. Received YYY; in original form ZZZ}

\pubyear{2024}

\begin{document}
\label{firstpage}
\pagerange{\pageref{firstpage}--\pageref{lastpage}}
\maketitle

\begin{abstract}
Stellar atmosphere modelling predicts the luminosity and temperature of a star, together with parameters such as the effective gravity and the metallicity, by reproducing the observed spectral energy distribution. Most observational data comes from photometric surveys, using a variety of passbands. We herein present the Python Stellar Spectral Energy Distribution (PySSED) routine, designed to combine photometry from disparate catalogues, fit the luminosity and temperature of stars, and determine departures from stellar atmosphere models such as infrared or ultraviolet excess. We detail the routine's operation, and present use cases on both individual stars, stellar populations, and wider regions of the sky. PySSED benefits from fully automated processing, allowing fitting of arbitrarily large datasets at the rate of a few seconds per star.
\end{abstract}

\begin{keywords}
Software -- stars: fundamental parameters -- Hertzsprung–Russell and colour–magnitude diagrams
\end{keywords}



\section{Introduction}

The Hertzsprung--Russell (H--R) diagram defines the fundamental plane through which stellar evolution is observed. The temperature ($T_{\rm eff}$) and luminosity ($L$) of stars have proved critical in many astrophysical domains, whether galactic archaeology \citep[e.g.][]{Ibata2014}, modelling stellar evolution \citep[e.g.][]{Bressan2012} or identifying the properties of exoplanet host stars \citep[e.g.][]{Torres2010}.

However, both the temperature and luminosity as used in the H--R diagram are hard to directly measure. Two primary solutions exist. The first method is to compute temperature and surface gravity ($\log g$) from spectra, and analyse stars in the $T_{\rm eff}$--$\log g$ diagram instead. This has the advantages of being independent of interstellar reddening and of probing physical conditions on the stellar surface, including returning abundances. However, taking spectra of sufficient quality for faint stars is impractical, $\log g$ is rarely returned with a precision much better than a factor of two, and computational overheads are substantial \citep[e.g.][]{Fouesneau2023}.

The second method is to rely on observational proxies, such as colour--magnitude or colour--colour diagrams. While instructive, using this two-dimensional description of stars' brightnesses in the era of multi-wavelength surveys only allows a limited subset of the available data to be used in any analysis, and lacks the ability to convert observations back to physical principles in the same way a true H--R diagram would. By analysing higher-dimensional colour--magnitude and colour--colour diagrams, i.e., by modelling the full spectral energy distribution (SED) of a star, we can incorporate all available information in our model fit.

This ability is particularly useful when identifying and characterising stars that are not well-fit by simple stellar models, for example, when different types of emitting or absorbing material in the star's environment cause infrared or ultraviolet excess in a star's SED compared to a stellar atmosphere model. Such material could be binary companions, gas discs or dust-laden stellar winds \citep[e.g.][]{Woods2011}. In identifying and modelling these objects properly, we can create catalogues of particular kinds of objects on a more physical basis. Using $T_{\rm eff}$ and $\log g$ from spectra, and $T_{\rm eff}$ and $L$ from SED fitting together can be useful in a variety of contexts, including fitting interstellar reddening and measuring stellar mass \citep[e.g.][]{McDonald2011}.

The concept of fitting stellar SEDs is not new. Many software packages have been defined to fit and analyse the SEDs of normal stars, dust-enshrouded stars and stellar populations \citep[e.g.][]{Ivezic1999,Robitaille2007,Gordon2016,Goldman2020}. However, these can be quite slow or inappropriate to compute large areas of sky, and require extinction-corrected SEDs to already have been computed which, in itself, can be a labour-intensive process. The widest-used system that both collects relevant data, processes and fits SEDs is the Virtual Observatory SED Analyser (VOSA) \citep{Bayo2008}. However, VOSA still relies on significant user input during the process of SED creation and fitting, and is impractical for analysing areas of the sky (e.g., a new field of observation with a wide-field telescope). There is a need for a fast tool that will automatically produce and analyse SEDs from both literature and new photometry, either for single sources or for a stellar population in an area of sky.

The software described here builds on the principles of the SED-fitting software outlined in \citet{McDonald2009}. Modified versions of this code were used in \citet{McDonald2012} to compute the SEDs of $\sim$120\,000 \emph{Hipparcos} stars, while \citet{McDonald2017} extended this analysis to $\sim$1.5 million \emph{Tycho--Gaia} cross-matched stars. This paper describes the public release of this code to form the Python Stellar Spectral Energy Distribution package ({\sc PySSED}). This new code represents a complete re-write of the original code into the Python language, and adds significant features, including automated access to databases via the Table Access Protocol, proper-motion-corrected cross-matching, distance-dependent extinction correction and automated outlier rejection of bad data during fitting.

The remainder of this paper is outlined as follows:
\begin{itemize}
    \item Section \ref{sec:Overview} describes the basic program architecture and functionality.
    \item Section \ref{sec:Examples} demonstrates the software on particular stars of note.
    \item Section \ref{sec:Applications} explores use cases for astrophysical environments.
    \item Section \ref{sec:Conclusions} concludes this work.
\end{itemize}


\section{Overview}
\label{sec:Overview}

\begin{figure}
 \includegraphics[width=\columnwidth, trim=0.cm 0.cm 0.cm 0.cm, clip]{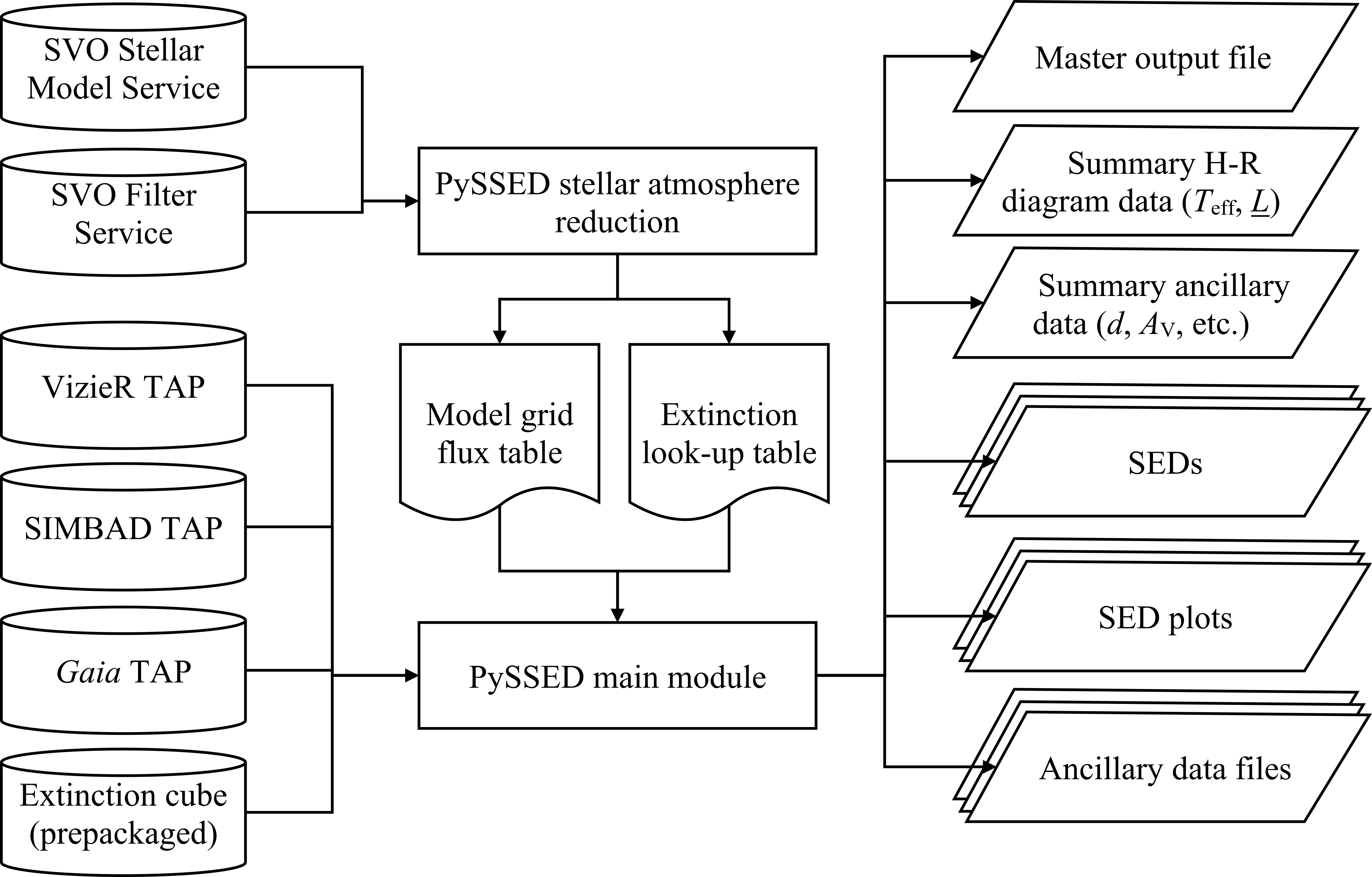}
 \caption{Flowchart showing the main data inputs and outputs of the {\sc PySSED} software.}
 \label{fig:flow1}
\end{figure}
\begin{figure}
 \includegraphics[width=\columnwidth, trim=0.cm 0.cm 0.cm 0.cm, clip]{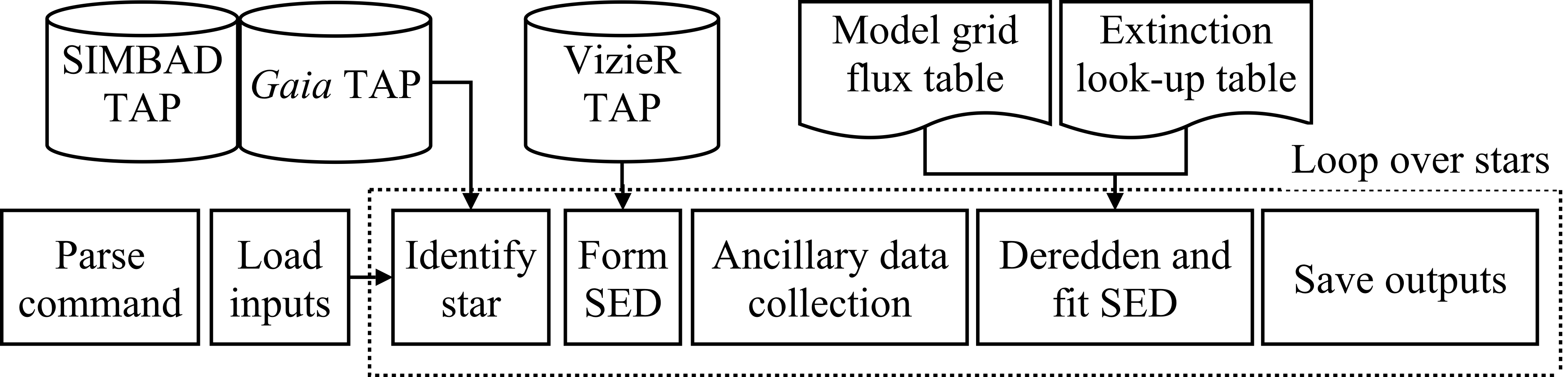}
 \includegraphics[width=\columnwidth, trim=0.cm 0.cm 0.cm 0.cm, clip]{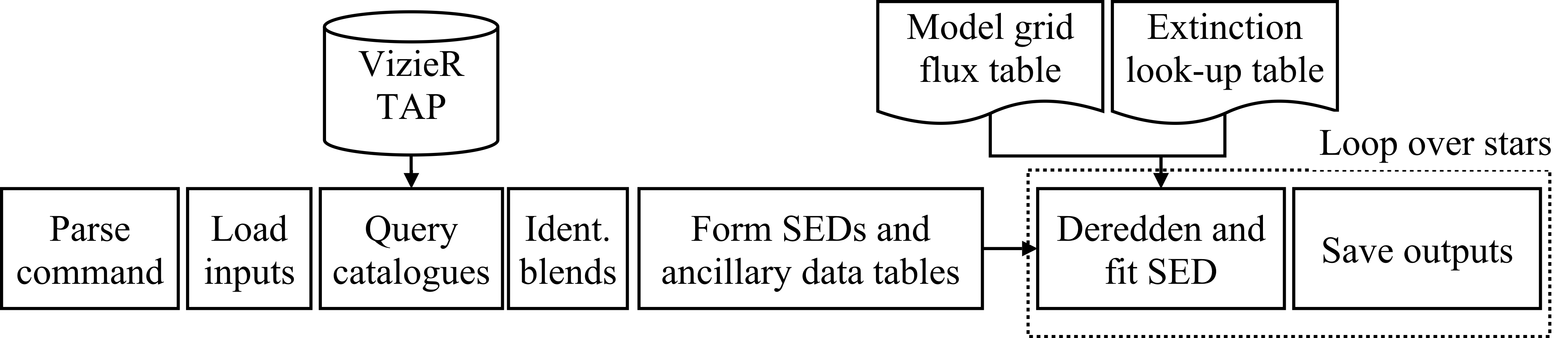}
 \caption{Flowchart for the {\sc PySSED} software in ``individual'' mode (top panel) and ``area'' mode (bottom panel).}
 \label{fig:flow2}
\end{figure}

{\sc PySSED} operates in two environments, either as a stand-alone command-line interface, or as a module that can be incorporated in a Python wrapper program.

The modular environment is designed to run remotely within the {\sc S-Phot} scientific data application (SDA) of the {\sc Explore} software suite\footnote{\url{https://explore-platform.eu/}} \citep{Cox2022}. This SDA benefits from a graphical user interface and its design emphasises interactions where individual data points and stars can be inspected in detail. Settings for data collection, collation, fitting and reporting can be changed via a user-uploaded setup file, but restrictions are placed on the sources and treatment of input catalogues for computational reasons. For many users who are interested in individual stars or simple fields, this will meet their needs. Example use-cases include teaching environments, quickly generate a plot of a star's SED for investigation, or generate a fit to an SED to identify a star's temperature and luminosity. These latter functions are particularly useful when approximate parameter estimates are needed \citep[e.g.][]{Decin2020}, or visually explore properties of individual stars \citep[e.g.][]{Driessen2020}.

The command-line environment is designed to be installed locally. This provides the opportunity to fine-tune settings for individual catalogues, filters and matching criteria, and add new catalogues, thus is designed to work better as part of larger projects. Example use cases include collation and cross-matching of large datasets \citep[e.g.][]{Ruffle2015}, analysis of stellar populations \citep[e.g.][]{Boyer2015}, derivation of higher-order parameters \citep[e.g., exoplanet habitable zones;][]{Chandler2016}, and astrometric image correction, which currently cannot easily be performed with existing tools unless with significant investment of user time.

There are two main modes of operation, designed to match to these corresponding environments. The ``individual'' mode operates on a single star or a list of such stars, given as either co-ordinates or names (to be resolved by {\sc simbad}\footnote{http://simbad.u-strasbg.fr/} \citep{Wenger2000}). Correspondingly, the ``area'' mode analyses all stars within an indicated region of the sky. The program's main data inputs and outputs are shown in Figure \ref{fig:flow1}, and broad flowcharts for the ``individual'' and ``area'' modes in Figure \ref{fig:flow2}. The remainder of this section describes the program execution.

\subsection{Setup}
\label{sec:setup}

The {\sc PySSED} package comes ready to use. However, an initial setup is required if the user wishes to add filters to the data, or change the underlying stellar model atmospheres. A setup program, ``{\sc makemodel}'', is provided for this purpose. {\sc Makemodel} convolves the stellar model atmospheres with the user-supplied list of filters and provides a look-up table, for each model atmosphere, containing the expected energy per square metre of emitting surface detectable through each filter (note that this implicitly assumes that data have been colour-corrected, which most large surveys have). A total power per square metre of emitting area is also recorded for each model, which is used to normalise the SED and compute the star's luminosity.

{\sc Makemodel} also assesses reddening. The model atmosphere is reddened by a standard amount (by default $A_V = 0.1, 3.1, 10$ and $31$ mag) using a \citet{Fitzpatrick1999} reddening law, and a ratio $A_{F,M}/A_V$ for each filter $F$ and each model $M$ is computed at each $A_V$. This provides an additional set of four look-up tables from which interstellar reddening can be corrected. (An option also exists in the main {\sc PySSED} to simply correct for reddening using the central wavelength of the filter, which results in approximately a 10 per cent increase in computation speed, at the expense of a $\lesssim$5 per cent error in temperature for moderate ($E-V \sim 1$ mag) extinction.)

An initial run of {\sc PySSED} will then interpolate these look-up tables onto a five-dimensional rectilinear grid in $T_{\rm eff}$, $\log g$, [Fe/H], [$\alpha$/Fe] and $A_V$, ready for the program to use.

\subsection{Collating SEDs and ancillary data}

\subsubsection{Data collection}

SEDs are collated on the basis of cross-match assignment to a master catalogue. By default, that master catalogue is \emph{Gaia} Data Release 3 \citep{GaiaDR3}, as it represents the largest set of stars with known distances (distances being needed to convert from apparent flux to extinction-corrected absolute flux). In the ``individual'' mode, the program begins a loop over all sources: if a set of co-ordinates is given, the closest \emph{Gaia} DR3 identifier will be selected, if it lies within a limiting search cone; if an identifier is given, the \emph{Gaia} DR3 identifier will be sourced from a {\sc simbad} cross-match, and a cone search on that position used if a cross-match doesn't already exist. The position and proper motion are extracted from the \emph{Gaia} data, and serve as the astrometric reference for cross-matching. In the ``area'' mode, all the \emph{Gaia} sources in the area (cone or box) will be extracted and the cross-matching takes place before the main loop over objects.

The program then loops over a user-supplied list of catalogues, broken into two parts: photometry and ancillary data. Each catalogue is assigned a small search cone that is used as a limiting allowed radius for cross-matches. For each source the proper motion of the master source is projected to the epoch of the catalogue, the closest catalogued match to the master source is then identified, and the match accepted if it falls within the radius of an expanded search cone that accounts for the uncertainty in the proper-motion correction. Default catalogues and their parameters are shown in Table \ref{tab:photosources}.

Data for individual (user-defined) filters are extracted, with filter properties taken from the Spanish Virtual Observatory Filter Profile Service (SVO FPS; \citealt{Rodrigo2012,Rodrigo2020})\footnote{http://svo2.cab.inta-csic.es/svo/theory/fps3/index.php}. Magnitudes are converted into fluxes and the basic SED is formed. Simultaneously, any non-photometric ancillary data requested by the user is extracted from the catalogue, such as parallax or distance information.

\begin{table*}
    \caption{Sources of photometry and their short-hand in this paper.}
    \label{tab:photosources}
    \centering
    \begin{tabular}{@{}l@{}l@{\ }l@{}c@{}c@{}c@{}c@{\ }c@{}l@{}}
        \hline\hline
        \multicolumn{1}{c}{Short-hand}    & \multicolumn{1}{c}{Full name}   & \multicolumn{1}{c}{VizieR}  & \multicolumn{3}{c}{Adopted} & \multicolumn{1}{c}{Filters}  & \multicolumn{1}{c}{\clap{Wavelength}}  & \multicolumn{1}{c}{Reference} \\
        \multicolumn{1}{c}{name}    & \multicolumn{1}{c}{\ }   & \multicolumn{1}{c}{table}  & \multicolumn{1}{c}{Epoch}  & \multicolumn{1}{c}{\clap{$r_{\rm beam}$}} & \multicolumn{1}{c}{\clap{$r_{\rm match}$}}  & \multicolumn{1}{c}{\ }  & \multicolumn{1}{c}{range}  & \multicolumn{1}{c}{\ } \\
        \multicolumn{1}{c}{\ }    & \multicolumn{1}{c}{\ }   & \multicolumn{1}{c}{\ }  & \multicolumn{1}{c}{(yr)}  & \multicolumn{1}{c}{($^{\prime\prime}$)} & \multicolumn{1}{c}{($^{\prime\prime}$)}  & \multicolumn{1}{c}{\ }  & \multicolumn{1}{c}{\ }  & \multicolumn{1}{c}{\ } \\
        \hline
        \emph{GALEX}        &  (Revised) all-sky survey of	& 	II/335/galex\_ais & 2008.5 & 2.8 & 1.5 & $FUV\,NUV$ & UV & \citet{Bianchi2017}\\
        \                   & \rlap{... Galaxy Evolution Explorer sources} &  &  &  &  &  & \\
        \emph{Gaia} DR3    & \emph{Gaia} Data Release 3  & I/355/gaiadr3    & 2016.0 & 0.5 & 0.4 & $B_{\rm P}\,G\,R_{\rm P}$    & Optical & \citet{GaiaDR3}\\
        \emph{Hipparcos}    & \emph{Hipparcos}: The New Reduction & I/311/hip2	& 1991.25 & 1.0 & 1.0 & $H_{\rm P}$  & Optical & \citet{vanLeeuwen07}\\
        \emph{Tycho}	    & Tycho                             & I/239/tyc\_main	& 1991.25 & 1.0 & 1.0 & $B_{\rm T}\,V_{\rm T}$ & Optical & \citet{Perryman97} \\
        Pan-STARRS          & Panoramic Survey Telescope \& & II/349/ps1 & 2011.9 & 0.5 & 0.5 & $grizy$ & Optical & \citet{Chambers16} \\
        \                   & \rlap{... Rapid Response System Data Release 1} &  &  &  &  &  & \\
        Morel               & ---                            & II/7A/catalog & 2000.0 & 5.0 & 5.0 & \llap{$U$}$BVRIJHKLM$\rlap{$N$} & Opt/IR & \citet{Morel1978} \\
        APASS       	    & American Association of  & II/336/apass9    & 2015.0	& 3.1	& 1.3   & $BVgri$ & Optical & \citet{Henden2015} \\
        \              	    & \rlap{... Variable Star Observers (AAVSO) Photometric All-Sky Survey} & \   & \	& \	& \   & \ & \ & \ \\
	    SDSS12         	    & Sloan Digital Sky Survey & V/147	    & 2008.0	& 1.4	& 0.5   & $u^{\prime}g^{\prime}r^{\prime}i^{\prime}z^{\prime}$ & Optical & \citet{SDSS} \\
        \              	    & (SDSS) Data Release 12 & \   & \	& \	& \   & \ & \ & \ \\
	    CMC15         	    & Carlsberg Meridian Catalog 15 & I/327/cmc15	    & 2005.25	& 1.5	& 0.5   & $r^\prime$ & Optical & $^1$ \\
	    2MASS	            & Two-Micron All-Sky Survey & II/246/out	    & 1999.5	& 1.0	& 1.0   & $JHK_{\rm s}$ & Near-IR & \citet{Cutri2003}\\
	    VVV	                & Visible and Infrared Survey & II/348/vvv2	    & 2016.0	& 0.3	& 1.0   & $ZYJHK_{\rm s}$ & Near-IR & \citet{Minniti2010}\\
        \              	    & \rlap{... Telescope for Astronomy (VISTA) Variables in the Via Lactea Data Release 2} & \   & \	& \	& \   & \ & \ & \ \\
	    VHS	                & VISTA Hemispheric Survey DR5  & II/367/vhs\_dr5	& 2016.0	& 0.3	& 1.0   & $YJHK_{\rm s}$ & Near-IR & \citet{McMahon2013}\\
	    All\emph{WISE}	    & $^3$ & II/328/allwise	& 2010.25	& 6.0	& 1.0   & [3.4]\,[4.6]\,[11.3]\,[22] & Mid-IR & \citet{Cutri2013}\\
	    cat\emph{WISE}	    & $^3$ & II/365/catwise	& 2010.25	& 6.0	& 1.0   & [3.4]\,[4.6]\,[11.3]\,[22] & Mid-IR & \citet{Marocco2021}\\
        un\emph{WISE}	    & $^3$ & II/363/unwise	& 2010.25	& 6.0	& 1.0   & [3.4]\,[4.6]\,[11.3]\,[22] & Mid-IR & \citet{Lang2014}\\
	    \emph{MSX}	        & \emph{Mid-course Space Experiment} & V/114/msx6\_main	& 1996.5	& 20.0	& 2.0   & $A\,B_1\,B_2$ & Mid-IR & $^4$\\
	    \emph{IRAS}	        & \emph{Infrared Astronomical Satellite} & II/125/main	    & 1985.0	& 180	& 15  & [12]\,[25]\,[60]\,[100] & Far-IR & \citet{Suarez06} \\
	    \emph{DIRBE}	    & \emph{Cosmic Background Explorer} & J/ApJS/154/673/DIRBE	& 1991.9	& 2520	& 10  & [1.25]--[240]$^5$ & IR & \citet{Smith2004} \\
        \              	    & \rlap{...  (\emph{COBE}) Diffuse Infrared Background Experiment} & \   & \	& \	& \   & \ & \ & \  \\
	    \emph{Akari} IRC    & \emph{Akari} Infrared Catalogue & II/297/irc	    & 2008.0	& 5.5	& 1.0   & [9]\,[18] & Mid-IR & \citet{Ishihara2010} \\
	    \emph{Akari} FIS    &  \emph{Akari} Far-Infrared Surveyor & II/298/fis	    & 2008.0	& 26.5	& 5.0   & [65]\,[90]\,[140]\,[160] & Far-IR & \citet{Kawada2007}\\
        \              	    & \rlap{... All-Sky Survey Bright Source Catalogue} & \   & \	& \	& \   & \ & \ & \  \\
	    \emph{Herschel}	    & \emph{Herschel} Photodetector Array & VIII/106/hppsc070& 2011.5	& 7.0	& 3.0   & [70] & Far-IR & $^6$\\
	    \ 	                & ...Camera and Spectrometer (PACS) & VIII/106/hppsc100& 2011.5	& 7.0	& 3.0   & [100] & Far-IR & $^6$\\
	    \ 	                & ... Point Source Catalogs & VIII/106/hppsc160& 2011.5	& 7.0	& 3.0   & [160] & Far-IR & $^6$\\
        \hline
        \multicolumn{9}{p{\textwidth}}{$^1$\url{http://svo2.cab.inta-csic.es/vocats/cmc15/docs/CMC15_Documentation.pdf} 
        $^2$\url{http://cds.u-strasbg.fr/denis.html} 
        $^3$All\emph{WISE}, cat\emph{WISE} and un\emph{WISE} represent different data reductions from the \emph{Wide-field Infrared Survey Explorer}. 
        $^4$\url{https://irsa.ipac.caltech.edu/Missions/msx.html} 
        $^5$ \emph{DIRBE} filters are [1.25], [2.2], [3.5], [4.9], [12], [25], [60], [100], [140] and [250].
        $^6$\url{https://doi.org/10.5270/esa-rw7rbo7}}\\
        \hline
    \end{tabular}
\end{table*}

\subsubsection{Data rejection and merging}

The data are then examined for user-defined rejection criteria. This can be used to mask bad data, including saturated or low-quality data, or to prefer one catalogue over another when multiple observations in nearby filters exist. The default criteria follow those outlined in the Appendix of \citet{McDonald2017}, and criteria can include data-warning flags or flux/magnitude cuts.

When more than one data point exists for a filter or for an ancillary data parameter, a merged value will be produced. This process differs slightly between photometric and ancillary data.

For photometric data, a substantial issue is that photometric errors on surveys are often under-estimated, since they only (normally) contain the photon noise associated with each measurement. However, when compiling an SED, the problems of blending, colour-correction, filter transmission errors and zero-point estimation must also be taken into account. For this reason, a user-supplied minimum fractional flux error is applied to each photometric data point. We estimate this at $\sim$10 per cent based on the typical scatter of well-fit stars (e.g., \citet{McDonald2017}, Figure B1). For the calculation of the merged flux, data points with a fractional error more than five times that of the most-precise observation are removed from consideration. The median flux of the remaining data points is calculated. If the individual data agree with this median with a certain threshold (defaulted to three times the observational error (3$\sigma$)), then they are averaged in quadrature; if no observations agree with this median, then the most precise value is assumed to be correct.

The process for ancillary data is largely similar but different thresholds can be set and merging is only attempted on numerical fields. A priority system is also set up so that observations from one catalogue can always be preferred over another, an example of which is described in the following section on distance estimation. Asymmetric upper and lower errors are allowed in ancillary parameters, however these are averaged to allow this procedure to proceed without overly complicating the processing, which can lead to undesirable effects.

\subsection{Distance estimation}

Distance is treated as an ancillary parameter, and is corrected in the same way as other ancillary data, with the exception that parallax measurements are inverted to provide distance estimates, then subsequently merged with other distance estimates as normal. However, distance is important as it is used to normalise the flux of the star to retrieve its luminosity. The treatment of distance is complex, as distances can come from a variety of physical methods, each of which is subject to biases on both individual objects and statistical samples.

The most important of these biases is the Lutz--Kelker bias in parallax measurements \citep{Lutz1973}, caused by the asymmetric probability tails generated when the parallax is inverted to provide a distance. Since, for reasons of computational speed, we only assign a simple error to data rather than considering the full probability distribution, this can cause severe biases in the luminosity estimation for individual objects. To combat this, we have prioritised distances from \citet{BailerJones2021} over distances obtained directly from \emph{Gaia} DR3 parallaxes. Bailer-Jones's distances probabilistically correct for the Lutz--Kelker bias: their geometric distances additionally correct for placement within a model of our Galaxy, while their photo-geometric distances additionally correct for the likelihood of a star of a particular colour being at a particular magnitude based on stellar evolutionary theory. These latter two corrections add their own biases in particular situations: for example, many \emph{Gaia} stars residing in external galaxies (e.g., the Magellanic Clouds or Sagittarius dwarf spheroidal) are mistakenly classed as Milky Way stars, while the photo-geometric distances can place stars undergoing unusual phases of evolution at incorrect distances. For balance, we choose their geometric distances.

However, many \emph{Gaia} stars do not have usable parallaxes, and many use-cases of SED-fitting codes may not wish to use \emph{Gaia} as a primary survey, and/or cover stellar populations at known distances (e.g., stellar clusters and Local Group galaxies). For these cases, we add another option: if there is no usable parallax or distance data, or if the uncertainty in that data exceeds a certain threshold, then a default distance can be used instead. In this way, when another galaxy is surveyed, its foreground stars and members can be separated based on parallax, and each given their correct properties.

\subsection{Extinction correction}

Once a distance is established, the photometry needs to be corrected for the extinction along the line of sight. {\sc PySSED} interfaces with another SDA of the {\sc explore} platform, {\sc G-Tomo}, which provides a line-of-sight extinction estimate based on a 3D cube derived from \emph{Gaia} data \citep{Lallement2022,Vergely2022}. This cube is supplied with the {\sc PySSED} package and covers a grid, centred on the Sun, which is 10\,kpc on each side and has a resolution of 50\,pc. The user can alternatively use the smaller (1200\,pc) cube from \cite{Gontcharov2017}.

As with distance estimation, there are circumstances where a single, fixed extinction is preferred, such as when working with stellar clusters. The option to set a default value is therefore also given.

\subsection{SED fitting}

\subsubsection{Model atmospheres}

Once a collated, extinction-corrected, distance-normalised SED has been created, it can be fit. While a blackbody fit is provided, it is expected that most analyses will fit to stellar atmosphere models instead. Any stellar model spectrum can be used, and a facility is provided to take models downloaded from the Spanish Virtual Observatory\footnote{\url{http://svo2.cab.inta-csic.es/theory/newov2/}} and reprocess them to be available in {\sc PySSED}. This pre-processing, which speeds up the fitting routine, is described in Section \ref{sec:setup}.

The default models used by {\sc PySSED} are the {\sc BT-Settl} AGSS2009 models \citep{Allard2011}, which are scaled to the \citet{Asplund2009} solar abundances. The {\sc BT-Settl} models are used because of their focus on cool stars and the comparative completeness of their grid in temperature, surface gravity and metallicity across the entire parameter range of stars. However, they do not adequately cover some combinations of stellar parameters, such as those of white dwarfs.

The process of fitting is detailed below, but can be summarised as follows. Once a distance is established, [Fe/H], [$\alpha$/Fe] and $A_V$ are established. An iterative process then takes place, whereby temperature is fit, then $\log(g)$ and the extinction correction revised, then the temperature is refit. Two iterations of this process have been found to yield stable results and, for speed, the first iteration uses a blackbody fit.

\subsubsection{Initial blackbody fitting}
\label{sec:bbfit}

The SED-fitting process requires the same four inputs as the stellar atmosphere grid: an estimated temperature, surface gravity, [Fe/H] and [$\alpha$/Fe].

To provide an estimated temperature, a blackbody is first fit using the Nelder--Mead method\footnote{ The Nelder--Mead method with a default initial simplex appears fairly optimal for almost all stars. For SEDs with a large number of outlying points, particularly those with extreme properties (e.g., white dwarfs or heavily self-extincted stars), the fit can become trapped in a local minimum or fail to converge (see the individual cases in this subsection for examples).} in the {\sc scipy.optimize.minimize} routine\footnote{\url{https://scipy.org/citing-scipy/}}. The minimising function is, at heart, a simple weighted reduced-$\chi^2$ minimisation, with the function to be minimised being
\begin{eqnarray}
\chi^2_{\rm r}(T) &=& \sum_{i=1}^{n} \left( \frac{\log F_{i,{\rm o}} - \log F_{i,{\rm m}}(T)}{n-1} \right) \frac{w_i}{\sum_{i=0}^n w_i} \\
w_i &=& \sqrt{\frac{\left(F_{i,{\rm o}} / \Delta F_{i,{\rm o}}\right)^2 + \Delta_{\rm offset}^2}{\Delta_{\rm offset}^2}}
        \ \Lambda \ w_{\rm prior}(T)\\
\Lambda &=& \exp\left(-\frac{(\log \lambda - \log \lambda_{\rm peak})^2}{\sigma_\Lambda^2}\right)^{p_\Lambda} .
\label{eq:chisq}
\end{eqnarray}
In this case, $F_{i,{\rm o}}$ is the observed flux in filter $i$ of $n$, and $\Delta F_{i,{\rm o}}$ its error. The corresponding model flux is $F_{\rm m}(T)$. Logarithms are used when evaluating $\chi^2$ to avoid bright points dominating the fit. The weight ($w_i$) is defined by the squared fractional errors, accounting for the aforementioned minimum photometric error, but also allowing an additional offset $\Delta_{\rm offset}$ to be applied (the default is 0.05). The factor $\Lambda$ is used to give further weight to the points close to the blackbody peak, which avoids problems in stars with mild ultraviolet or infrared excess, and is controlled by width and severity parameters that default to $\sigma_\Lambda = 0.5$ and $p_\Lambda = 2$, respectively. Fitting is also limited to a pre-determined wavelength range (by default $\lambda = 0.1-20\,\mu{\rm m}$) for similar reasons.

The weight can additionally be modified by a prior weight ($w_{\rm prior}(T)$) if a prior temperature constraint exists. The option exists to read in a spectroscopic temperature as an ancillary parameter, and use it as a prior constraint. Mathematically, the prior constraint consists of two error functions, one inverted, which are multiplied together. In this way, a single temperature with an error can be applied, or a range of possible temperatures. By default, the upper temperature limit is set to 200\,000\,K, with a characteristic width to the error function of 10\,000\,K, while the lower temperature limit is set to 10\,K with a width of 1\,K, which conservatively cover the expected range for physical stars, including allowing fits for cooler, obscured objects.

Once complete, the best-fitting temperature is returned and the normalisation of the blackbody fit used to determine the stellar bolometric luminosity.

\subsubsection{Stellar atmosphere model fitting}

Once a blackbody fit has provided a starting estimate for the stellar temperature, the surface gravity, [Fe/H] and [$\alpha$/Fe] can be estimated.

By default, to calculate [Fe/H] or [$\alpha$/Fe], a scaling law is applied where ${\rm [Fe/H]} = -\arctan(Z)$ and $[\alpha/{\rm Fe}] = \arctan(Z)/5$, where $Z$ is the height above the Galactic Plane in kpc \citep[cf., e.g.][]{GaiaDR3Chem}. Alternatively, the user can specify values of [Fe/H] and [$\alpha$/Fe] (and $E(B-V)$ if they wish). These user-supplied values can be entered in the setup file, which applies them to the analysis as a whole (e.g., for work in stellar clusters). Values for individual stars can be loaded as ancillary data files (e.g., to be used when spectral fits are available).

If a mass is assumed, the surface gravity can then be estimated from the temperature and luminosity estimated from the blackbody fit. The precise mass chosen has only a second-order effect on the resulting fit. For temperature and luminosities consistent with a main-sequence star, the mass is estimated as $M \propto L^{1/3.5}$. For giant stars, a core mass is estimated using $M_{\rm core}[\textrm{M}_\odot] = L / (62\,000 \textrm{L}_\odot) + 0.487$ \citep{Blocker1993}, an initial mass is estimated from this using $M_{\rm init}[\textrm{M}_\odot] = (M_{\rm core} - 0.3569) / 0.1197$ \citep{Casewell2009}, and (since mass-loss rate increases during AGB evolution) it is assumed that $M = (2 M_{\rm init} + M_{\rm core}) / 3$. This provides a minimum mass of 0.69 M$_\odot$ for the lowest-luminosity giants which, given the low impact of surface gravity on our fits, is tolerably close to the true value.

These estimates are sufficient for estimating the mass of main-sequence stars and most giant-branch stars with tolerable accuracy (since surface gravity only has a second-order effect on the other parameters). However, these mass estimates will deviate significantly in some cases, such as hot horizontal-branch stars, white dwarfs, and stars out of thermal equilibrium.

Once all four starting parameters are estimated, the fitting process is repeated, replacing the blackbody fluxes with the filter-convolved stellar atmosphere models as $F_{i,{\rm m}}$.

\subsubsection{Iterative fitting and outlier rejection}

Once a stellar atmosphere model is fit, it can be inspected for outliers. Outliers may occur for numerous reasons, including: bad cross-matching of stars between surveys, blended stellar light in surveys of differing angular resolutions, poor-quality data for surveys of differing sensitivity to bright/faint objects and astrophysical sources of variability.

The SED is first inspected for points that are not in agreement with the stellar atmosphere model. If the ratio of observed to modelled flux exceeds a certain value (by default $F_{i,{\rm o}}/F_{i,{\rm m}} < 0.8$ or $> 1.25$) then the point is classified as a potential outlier.

If outliers are found, the worst-fitting point is removed and the fitting process re-run. If the $\chi^2_{\rm r}$ is reduced by more than a threshold factor (by default 1.5), then rejecting that point is deemed to improve the fit, the point is masked in the SED, and the outlier process run again.

If removing a single potential outlier does not sufficiently improve the $\chi^2_{\rm r}$, then the process is re-run, removing two outliers. This is necessary because of correlations between the filters (e.g., an offset that affects an entire multi-wavelength catalogue). In this case, the necessary reduction in $\chi^2_{\rm r}$ is squared (thus, by default, a factor of $1.5^2 = 2.25$). The process continues with three points, then four points, etc., until either all potential outliers have been tested, or a maximum sequential number of outliers (by default eight) have been examined together. The process is also designed to stop when a maximum number of outliers has been removed (by default 15), when the remaining SED contains a minimum number of points (by default 5), or when more than a specified fraction of points has been removed from the SED (by default half).

Once outlier rejection is complete, the surface gravity is refined as before, and a final model and parameter set are returned.

\subsection{Outputs and benchmarks}

All outputs are returned by default, although the option to suppress each output exists. For individual stars, files are returned contain the SED in tabular format, the SED in graphical format and the ancillary data. A separate file lists the returned temperature and luminosity for all individual targets, or targets in a list. These outputs are designed for small datasets. For larger datasets, the SEDs (including cross-identifications and fitted parameters) are also returned in a single, master table.

For area searches, additional graphs can be generated for analysis and for interrogation of input data quality and consistency. These include:
\begin{itemize}
    \item an H--R diagram;
    \item one- and two-dimensional plots of pointing offsets to diagnose astrometry issues;
    \item a plot of proper motions to identify clustering;
    \item plots in each filter of excess flux (i.e., $F_{i,{\rm o}}/F_{i,{\rm m}}$) as a function of catalogue flux, to diagnose issues with saturation or poor signal-to-noise;
    \item plots in each filter of excess flux as a function of spatial position, to diagnose issues with intra-catalogue variation, such as background light; and
    \item a matrix plot of excess flux in each filter, to diagnose cross-catalogue issues, including errors in filter transmission functions and photometric zero-points.
\end{itemize}

Benchmarks were calculated using a four-core AMD Ryzen 5 2500U CPU with 32 GB of RAM. In the default setup, an star can (on average) be fully modelled in 28 seconds, requiring 5\,s to load the relevant Python modules, 20\,s to query the servers for data, and 3\,s to model the star. The rate-limiting step is therefore in the data-server queries. These also apply to searches of lists of stars, since the queries need repeated for each source, meaning ``list'' searches with the default setup take $\sim$23 seconds per star to run.

Conversely, searches in the ``area'' mode need only one query per catalogue, meaning these can be performed much faster. Execution time depends on the number of sources passed for fitting (as many can be rejected due to lack of data or because of user-defined cuts). However, as an example, the case of NGC\,104 presented later in the paper queried 45\,956 \emph{Gaia} objects and fitted 16\,062 of these. This process took 12.5 hours wall time (2.8 seconds per star) of which only the first half hour was used to query databases and collate the data into SEDs of individual objects. Typical CPU usage during this time was 2.0 cores: while {\sc PySSED} is not natively parallelised, multiple instances can be run simultaneously in different directories to speed up fitting larger areas. This ability to rapidly generate SED fits over significant areas of the sky demonstrates the substantial benefits of {\sc PySSED} over existing SED-fitting software.


\section{Tested examples}
\label{sec:Examples}

We anticipate a major use case for {\sc PySSED} is the quick analysis of individual stars' SEDs, and to approximate their potential for follow-up study (e.g., to select targets for spectroscopy). In this section, we identify the successes and limitations of {\sc PySSED} when it is applied to selected problem cases. The section closes with a statistical sample across a range of temperatures.

\subsection{Nearby and high proper-motion stars}

\begin{figure}
 \includegraphics[width=\columnwidth, trim=0.39cm 0.39cm 0.39cm 0.39cm, clip]{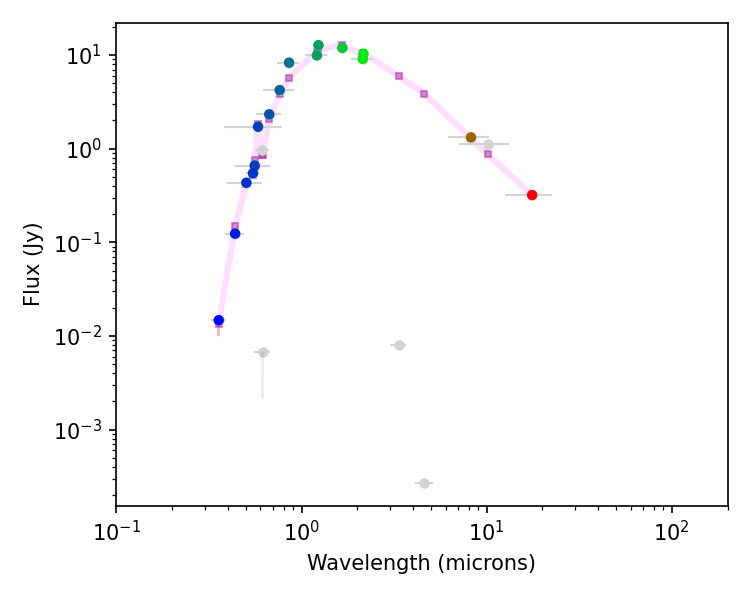} 
 \caption{The SED of Barnard's star. Points with colours show photometric bands used in the final fit. Lighter, grey points indicate photometric data masked from the final fit. Horizontal bars show the effective width of each filter; vertical error bars show the photometric uncertainty. The purple squares and connecting purple line show the best-fitting model atmosphere.}
 \label{fig:Barnard}
\end{figure}

\begin{table}
 \caption{Derived and published ``gold-standard'' properties of Barnard's star.}
 \label{tab:Barnard}
 \begin{tabular}{lcclc}
  \hline
  Property & {\sc PySSED} & \citet{Pineda2021} & Unit & Error (per cent)\\
  \hline
  $T_{\rm eff}$ & 3143    & $3223 \pm 17$                   & K & $2.5 \pm 0.5$\\
  $L$           & 0.00328 & $0.00339^{+0.00006}_{-0.00007}$ & L$_\odot$ & $3.2 \pm 1.8$\\
  $R$           & 0.193   & $0.187 \pm 0.001$               & R$_\odot$ & $3.1 \pm 0.5$\\
  \hline
 \end{tabular}
\end{table}

\begin{figure}
 \includegraphics[width=\columnwidth, trim=0.39cm 0.39cm 0.39cm 0.39cm, clip]{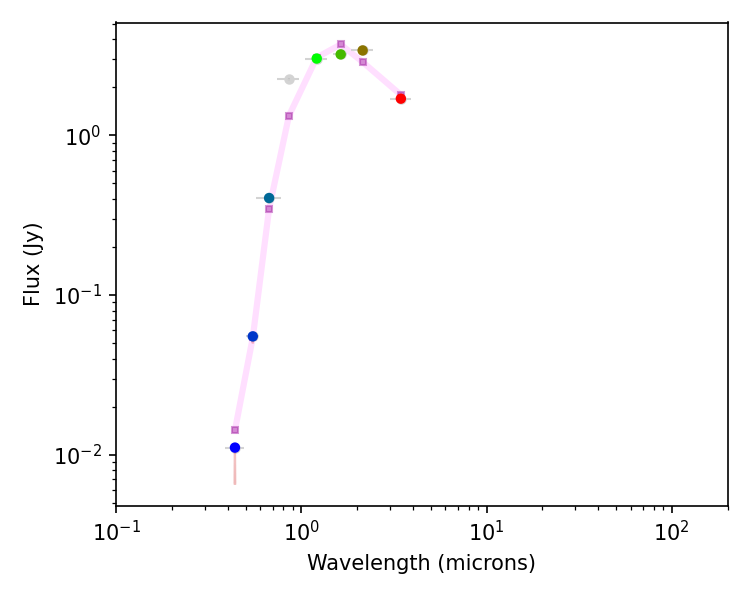}
 \caption{The SED of UV Ceti. Points and lines as in Figure \ref{fig:Barnard}.}
 \label{fig:UVCeti}
\end{figure}

Correcting for high proper motion can be difficult, as the uncertainties involved in the proper-motion correction can be large, and the resulting search cone can easily include confusing stars. Imprecision in the proper motion and (for the very closest stars) movement of the star in its parallactic arc can lead to some cross-identifiers being missed.

As a critical test of the tool's ability to obtain the SEDs of high-proper-motion stars, we try to reconstruct the SED of the highest-proper-motion star, Barnard's star, using {\sc PySSED}'s default setup. This star has a very high proper motion of approximately 10 arcsec/yr. The {\sc simbad} co-ordinates for Barnard's star do not resolve to the \emph{Gaia} DR3 source, so {\sc PySSED} defaults to the corresponding \emph{Hipparcos} source. Cross-matches were obtained from \emph{Hipparcos} back to \emph{Gaia} DR3, and also to the Pan-STARRS, \citet{Morel1978}, CMC15, 2MASS, both \emph{AllWISE} and \emph{catWISE}, \emph{Akari} and \emph{IRAS}.

Barnard's star is heavily saturated in the \emph{WISE} data, leading to a dramatic under-prediction of the flux in the \emph{AllWISE} catalogue (the two outlying infrared points with low fluxes in Figure \ref{fig:Barnard}). As with all \emph{WISE} objects with Vega magnitudes $<$5, the 3.4 and 4.6 $\mu$m data from \emph{catWISE} are automatically selected in preference, as this catalogue deals better with these saturated data.

Initial blackbody fitting identified a temperature of 2621\,K, following which a mass of 0.176\,M$_\odot$ and $\log(g)$ of 4.96 were estimated. Fitting with a stellar model atmosphere revised this to 2802\,K. The Pan-STARRS $r$-band data is a clear outlier (the shortest-wavelength sub-luminous outlier of the three in Figure \ref{fig:Barnard}), which occurs because several data artefacts surround Barnard's star in the Pan-STARRS catalogue and one of these was chosen instead of the correct cross-match. Rejecting this outlier improves the goodness-of-fit by a factor of 60. The $I$-band measurement of \citet{Morel1978} was also identified as an outlier: this catalogue provides a key constraint on the fluxes of bright stars subject to saturation in the major surveys, but the nature of its compilation means that the heterogeneous filter transmission profiles (especially in $I$-band) can lead to discrepancies such as this. Removing this outlier improves the goodness-of-fit by a factor of two. No other outliers were identified.

{\sc PySSED} returned a final temperature of $T_{\rm eff} = 3143\,{\rm K}$, luminosity of $L = 0.00328\,{\rm L}_\odot$ and radius of $R = 0.193\,{\rm R}_\odot$.  These results are compared to ``gold-standard'' published estimates from \citet[][(GJ 699 in their paper)]{Pineda2021} in Table \ref{tab:Barnard}. The errors displayed in this table of $\sim$3 per cent in fitted $T_{\rm eff}$, $L$ and $R$ are fairly typical of {\sc PySSED}'s performance for late-type stars with good photometric coverage.

In general, high-proper-motion stars are well-fit by {\sc PySSED}, provided there is an accurate proper-motion solution in the star's \emph{Gaia} or \emph{Hipparcos} counterparts. Many nearby stars lack such solutions, so cannot be fit by {\sc PySSED}. Often, this is either because they are very late-type stars that are too faint to have counterparts, or are double stars (discussed below), though there are stars (e.g., Wolf 359) that are bright enough to be observed by \emph{Gaia} but lack any record in \emph{Gaia} DR3. There are also a small number of stars (e.g., Proxima Centauri) where either the {\sc simbad} position does not map to the \emph{Gaia} DR3 object, and/or where the \emph{Gaia} DR3 proper motion does not adequately project back to the source position at different epochs. In these cases, sometimes specifically requesting fitting of the \emph{Gaia} counterpart produces a different fit. This fit is not necessarily an improvement. An example is the binary UV Ceti + BL Ceti. Requesting a fit of UV Ceti itself produces a reasonably accurate fit of 2881\,K and 0.00193\,L$_\odot$ using only photometry from \citet{Morel1978} (cf., 2784 and 2728\,K, and 0.00147 and 0.00125\,L$_\odot$; \citealt{MacDonald2018}; Figure \ref{fig:UVCeti}). Specifically requesting a search on the \emph{Gaia} counterpart, Gaia DR3 5140693571158946048, also identifies the \emph{Gaia}, PanSTARRS, \emph{GALEX}, CMC15, VHS and \emph{IRAS} cross-matches. However, the poor-quality photometry of this comparatively bright star, and UV excess in the \emph{GALEX} data (possibly due to chromospheric activity), cause a poorer-quality fit of 2491\,K and 0.00230\,L$_\odot$. 

\subsection{Double stars}

Double stars present a particular problem for an automated SED fitter, as the two stars will appear merged in some bands, but not in others. In double stars near to the Earth, additional problems of proper-motion acceleration become important as the stars move in their orbits: while $\alpha$ Cen A and B are correctly identified by {\sc PySSED} using their \emph{Hipparcos} identifiers, the resulting proper-motion vectors do not align with the stars' current positions. In general, this effect only appears significant in the closest systems, such as $\alpha$ Cen and 61 Cygni.

Trials were performed on stars extracted from the 2020 edition of the Washington Visual Double Star Catalogue \citep{Mason2001}. Since the individual components are not listed, this test was performed by manual identification of each component on images. Double stars were selected where the brighter component has a magnitude of 9.0 $\pm$ 0.2 mag, and stars of different separations were tested for blending or incorrect cross-matches. Tests were run with both equal-brightness companions, and companions that were 2.0 $\pm$ 0.2 mag fainter. In both cases, good SEDs were formed for most double stars down to the limit where photometry became merged in the source catalogues. In the default surveys, this is about 3$^{\prime\prime}$ for stars of equal brightness, but this radius increases as fainter companions become lost in the glare of the brighter star. At closer separations, both the quality of the SEDs and the likelihood that the stars have a retrievable distance declines.

\subsection{Stars with circumstellar material, background light or variability}

\begin{figure}
 \includegraphics[width=\columnwidth, trim=0.39cm 0.39cm 0.39cm 0.39cm, clip]{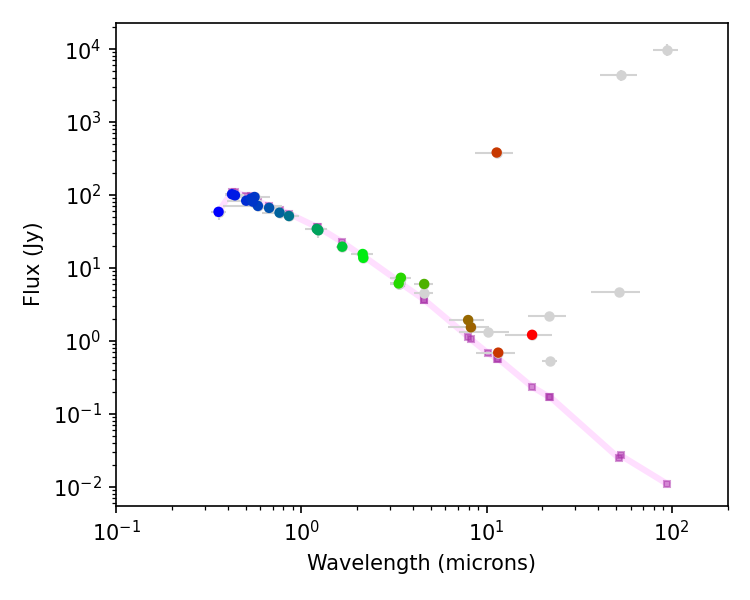}
 \caption{The SED of Merope. Points and lines are as Figure \ref{fig:Barnard}.}
 \label{fig:Merope}
\end{figure}

\begin{table}
 \caption{Derived and published ``gold-standard'' properties of Merope.}
 \label{tab:Merope}
 \begin{tabular}{lcclc}
  \hline
  Property & {\sc PySSED} & \citet{Zorec2016} & Unit & Error (per cent)\\
  \hline
  $T_{\rm eff}$ & 11\,009 & $14\,550 \pm 310$      & K         & $24.3 \pm 2.1$\\
  $L$           & 423     & $927^{+46}_{-44}$      & L$_\odot$ & $54.4 \pm 5.0$\\
  $R$           & 5.66    & $4.80^{+0.33}_{-0.31}$ & R$_\odot$ & $17.9 \pm 6.9$\\
  \hline
 \end{tabular}
\end{table}

The default extinction correction algorithm (based on 3D cubes of Milky Way extinction) does not take into account material in the local environment of individual stars. Emitting, absorbing or reflecting material in the same line of sight as the target star can therefore cause problems with the SED fit. This includes young stars in nebulae and mass-losing AGB stars. It is impossible to accurately remove reflection nebulae from a compiled SED, thus we rely instead on having the correct input photometry. However, catalogue photometry for stars in nebula can be very wrong, especially if the aperture is large.

This is exemplified in Figure \ref{fig:Merope}, which shows the Pleiades star Merope (23 Tau). The 12, 25 and 60 $\mu$ photometry from the \emph{IRAS} satellite lie well above the nominal SED, while \emph{WISE} and \emph{Akari} infrared photometry lie above the SED to a lesser extent, commensurate with the different beam sizes of these missions. As is typical with hot stars, the optical part of the SED is poorly constrained, but the SED is also made slightly redder and under-luminous because of the reflected light and extra (unaccounted-for) nebular absorption. While this can cause similar stars to be modelled as too cool and too small, the model for Merope in particular is too hot and too large (Table \ref{tab:Merope}).

\begin{figure}
 \includegraphics[width=0.91\columnwidth, trim=0.00cm 0.95cm 0.00cm 0.39cm, clip]{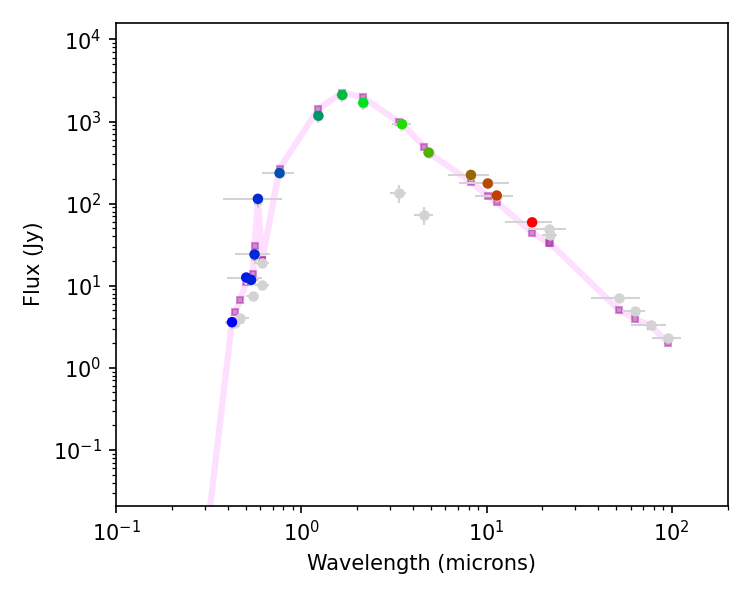}
 \includegraphics[width=0.91\columnwidth, trim=0.00cm 0.95cm 0.00cm 0.39cm, clip]{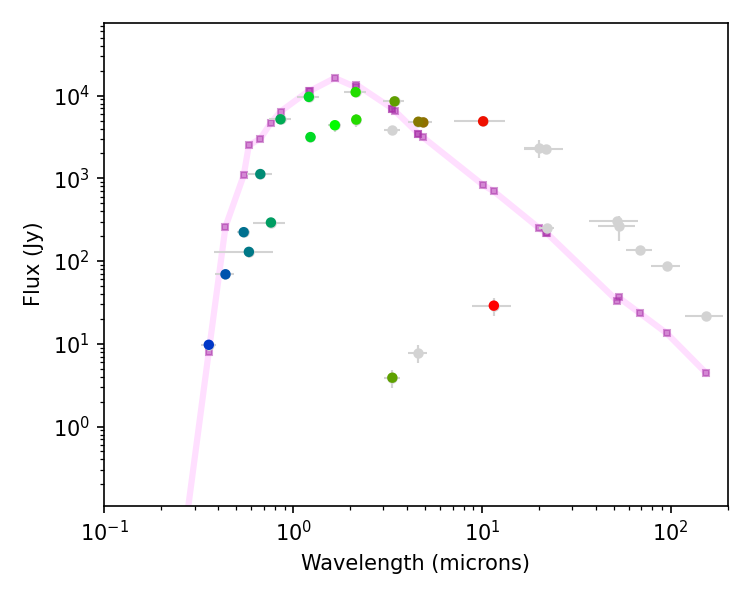}
 \includegraphics[width=0.91\columnwidth, trim=0.00cm 0.95cm 0.00cm 0.39cm, clip]{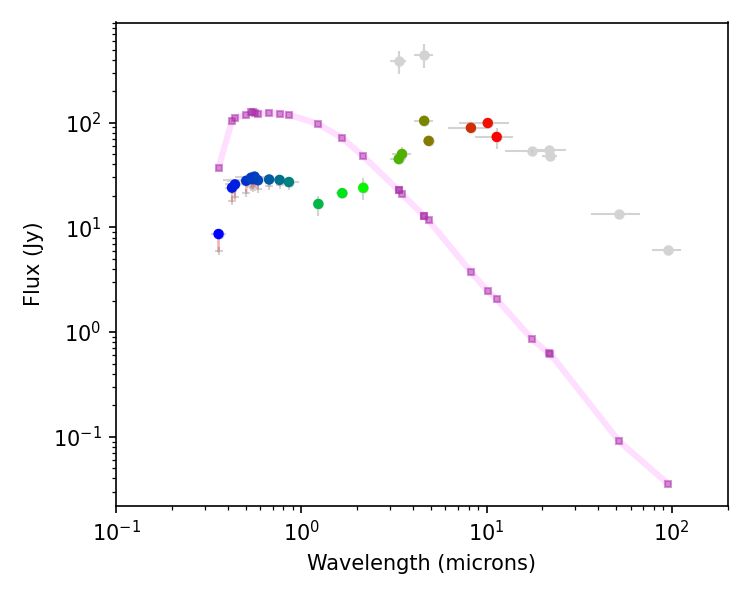}
 \includegraphics[width=0.91\columnwidth, trim=0.00cm 0.39cm 0.00cm 0.39cm, clip]{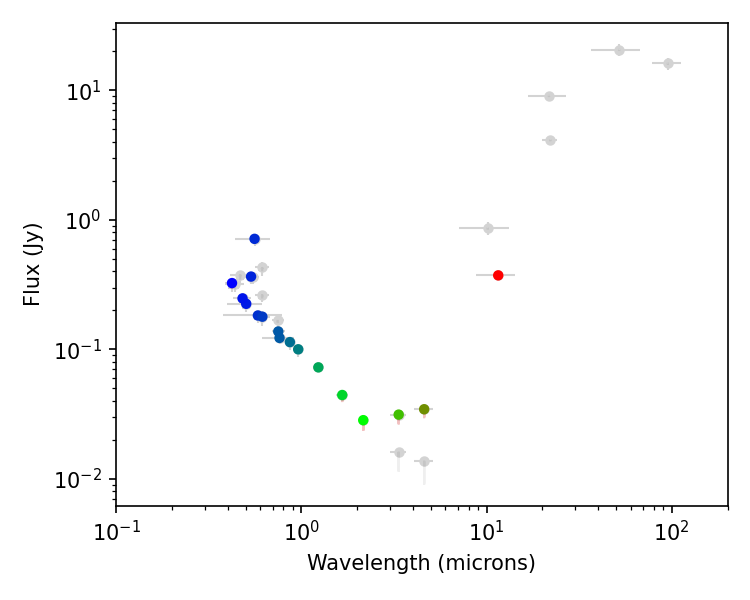}
 \caption{The SEDs of evolved stars with progressively later evolutionary states. From top to bottom, the SEDs of the AGB stars EU Del and Mira, the post-AGB star 89 Her and the planetary nebula NGC\,2392. Points and lines are as Figure \ref{fig:Barnard}.}
 \label{fig:AGB}
\end{figure}

Circumstellar material is easier to identify, as this usually leads to an infrared excess in the star's SED, and SED fitting is a powerful way to identify this. Outlier rejection can remove this infrared excess if it is sufficiently distinct in the SED. However, circumstellar material tends to also lie in front of the star, thus absorbs some of its light. This reduces the effective photometric temperature of the star, as derived from the integrated SED, and causes the temperature derived from the SED fit to depart from temperatures derived from spectra. In cases where the infrared excess represents a substantial fraction of the stellar flux, {\sc PySSED} may either fit the infrared excess as a blackbody, or fail to fit the star entirely. Figure \ref{fig:AGB} shows the progression from naked AGB star through to planetary nebula, and the effect this has on both the SED and its fit. The fit to NGC\,2392 failed due to the extreme nature of the source.

Stars with circumstellar material also tend to be variable stars. This scatters a star's photometry compared to the model SED, causing difficulty in outlier detection and SED fitting. By using time-averaged photometry, such as \emph{Gaia}'s, it is possible to reduce this scatter, but it contributes to additional uncertainty in the fit.

\subsection{Very hot and very cool stars}

\begin{figure}
 \includegraphics[width=\columnwidth, trim=0.39cm 0.39cm 0.39cm 0.39cm, clip]{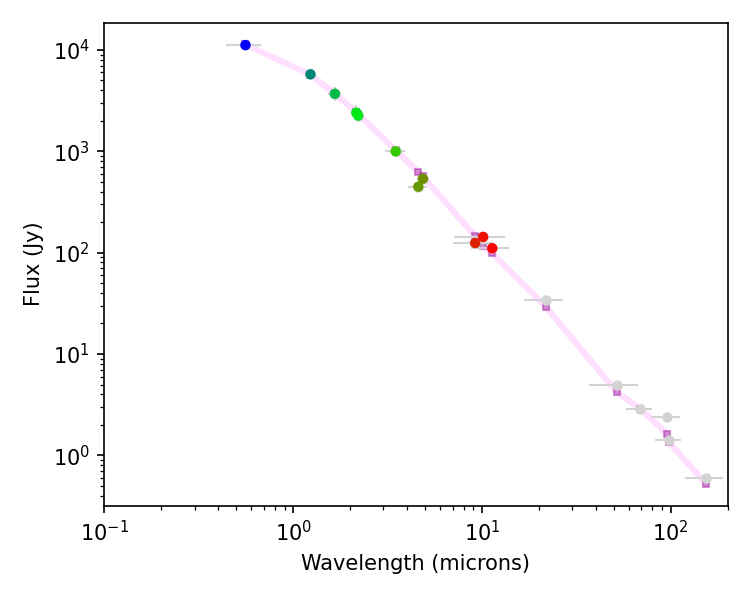}
 \caption{The SED of Sirius. Points and lines are as Figure \ref{fig:Barnard}.}
 \label{fig:Sirius}
\end{figure}

\begin{table}
 \caption{Derived and published ``gold-standard'' properties of Sirius A.}
 \label{tab:Sirius}
 \begin{tabular}{lcclc}
  \hline
  Property & {\sc PySSED} & \citet{Bond2017} & Unit & Error (per cent)\\
  \hline
  $T_{\rm eff}$ & 8799    & $9845 \pm 64$        & K         & $11.8 \pm 0.7$\\
  $L$           & 18.38   & $24.74 \pm 0.70$     & L$_\odot$ & $34.6 \pm 2.8$\\
  $R$           & 1.8557  & $1.7144 \pm 0.0090$  & R$_\odot$ & $7.6 \pm 0.5$\\
  \hline
 \end{tabular}
\end{table}

\begin{figure}
 \includegraphics[width=\columnwidth, trim=0.39cm 0.39cm 0.39cm 0.39cm, clip]{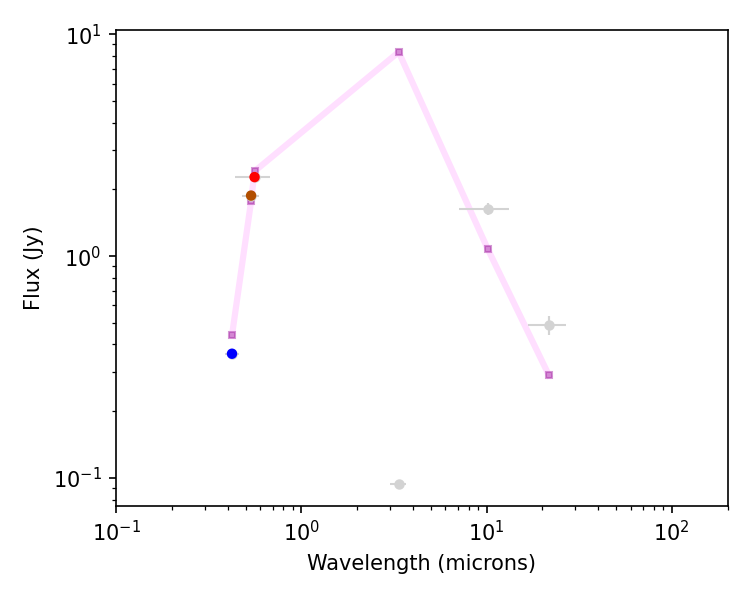}
 \caption{The SED of GJ 15 (GX + GQ And). Points and lines are as Figure \ref{fig:Barnard}.}
 \label{fig:GJ15}
\end{figure}

\begin{table}
 \caption{Derived and published ``gold-standard'' properties of GJ 15.}
 \label{tab:GJ15}
 \begin{tabular}{lcclc}
  \hline
  Property & {\sc PySSED} & \citet{Pineda2021} & Unit & Error (per cent)\\
  \hline
  $T_{\rm eff}$ & 3600    & $3601^{+12}_{-11}$        & K         & $0.0 \pm 0.3$\\
  $L$           & 0.0219  & $0.0224 \pm 0.0002$       & L$_\odot$ & $2.3 \pm 1.0$\\
  $R$           & 0.381   & $0.393^{+0.009}_{-0.008}$ & R$_\odot$ & $3.1 \pm 2.3$\\
  \hline
 \end{tabular}
\end{table}

The SED fitting for very hot stars fails when there is insufficient constraint on the UV or optical parts of the SED, or when the reddening correction is over-estimated. This leads to an increasingly unconstrained temperature, for example in the extreme example of Sirius A (Figure \ref{fig:Sirius}; Table \ref{tab:Sirius}), where the optical photometry relies entirely on the \emph{Hipparcos} data (the SIMBAD cross-match of Sirius to \emph{Gaia} DR3 records Sirius as an eighth magnitude star: this is apparently a bad cross-match to Sirius B). Sirius B is missed because it lacks its own photometry in major surveys. Stars hotter than Sirius exhibit progressively worse fits due to lack of photometric constraint to the stars' Wien tails.

For very cool stars, the primary difficulty is normally a lack of optical information to provide a parallax (e.g., Luhman 16 is unmodellable in the default settings as it lacks a \emph{Gaia} parallax). However, this can be overcome by specifying the star's distance in the setup file. A second difficulty is the lack of photometry to provide constraints. An example of this latter problem is GJ 15 (Figure \ref{fig:GJ15}; Table \ref{tab:GJ15}) though, in this case and in general, good-quality fits can still found with relatively sparse data, providing that the data is accurate.

\subsection{Faint stars}

As \emph{Gaia} is relatively deep for an all-sky survey, the majority of stars in \emph{Gaia} DR3 are relatively faint, and often lack the multi-wavelength coverage needed to properly constrain the SED. This can lead to substantial scatter in the H--R diagram. The example of GJ 15, above, is a case in point. In cases where only high-precision photometry is desired, the option exists to limit the accepted data used in the SEDs and in the fitting. These limits can either be as a strict magnitude limit for an individual filter, a percentage error in the observation, or an arbitrary cutoff based on another variable in that survey.

\subsection{Crowded regions}

Dense star fields cause many different problems in photometric surveys (see also Section \ref{sec:Globs}). One of the most important for cross-matching is the lack of consistency in the stars that are recorded. In lower-resolution surveys, one star will be detected while, in higher-resolution surveys, multiple stars may be. This generally causes steps or offsets in the SED and can lead to a larger fraction of outliers. Additionally, distances from parallaxes and proper-motion corrections to astrometry may be of poorer quality in crowded regions.

\begin{figure}
 \includegraphics[width=\columnwidth, trim=0 0.5cm 0 1.2cm, clip]{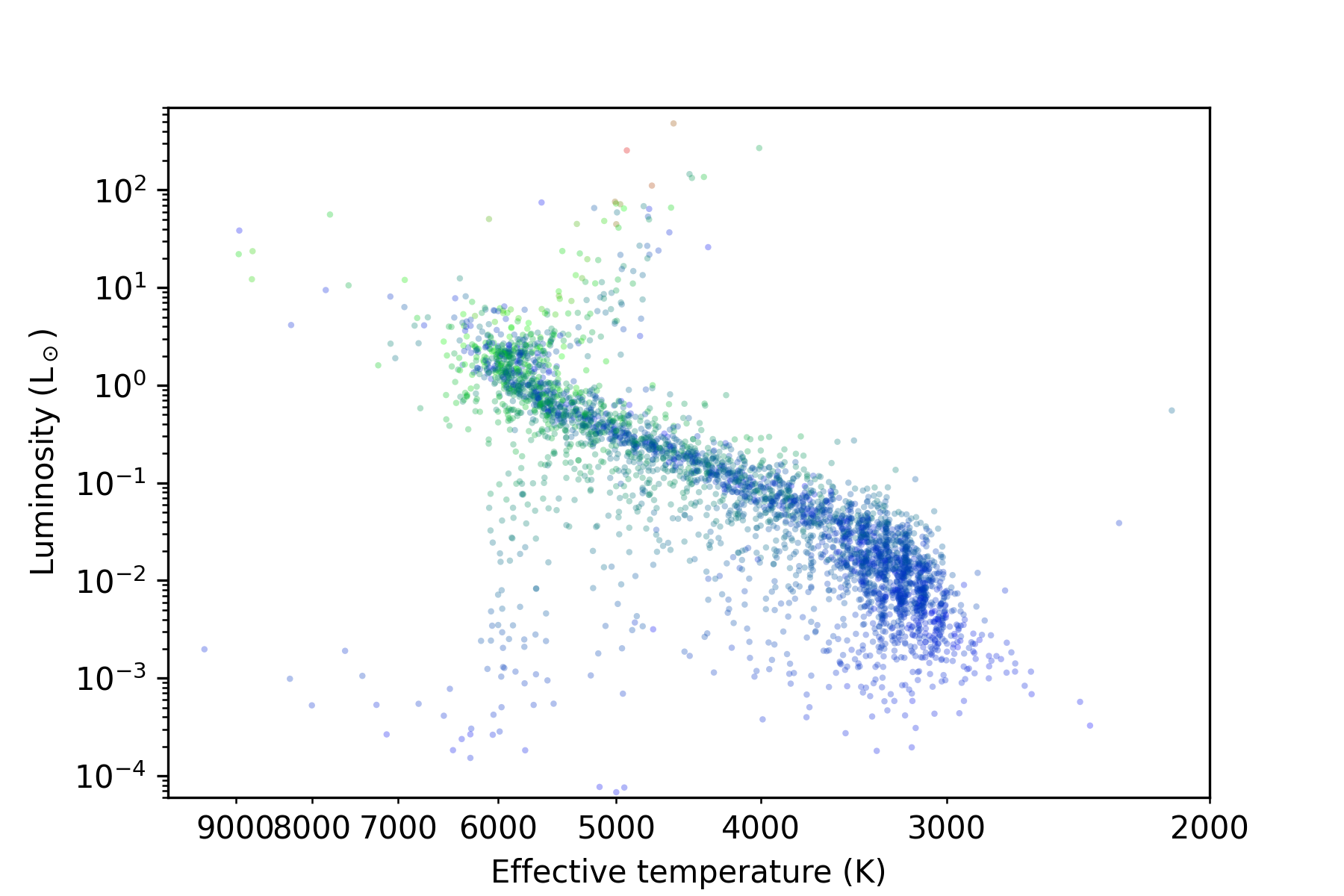}
 \includegraphics[width=\columnwidth, trim=0 0cm 0 1.2cm, clip]{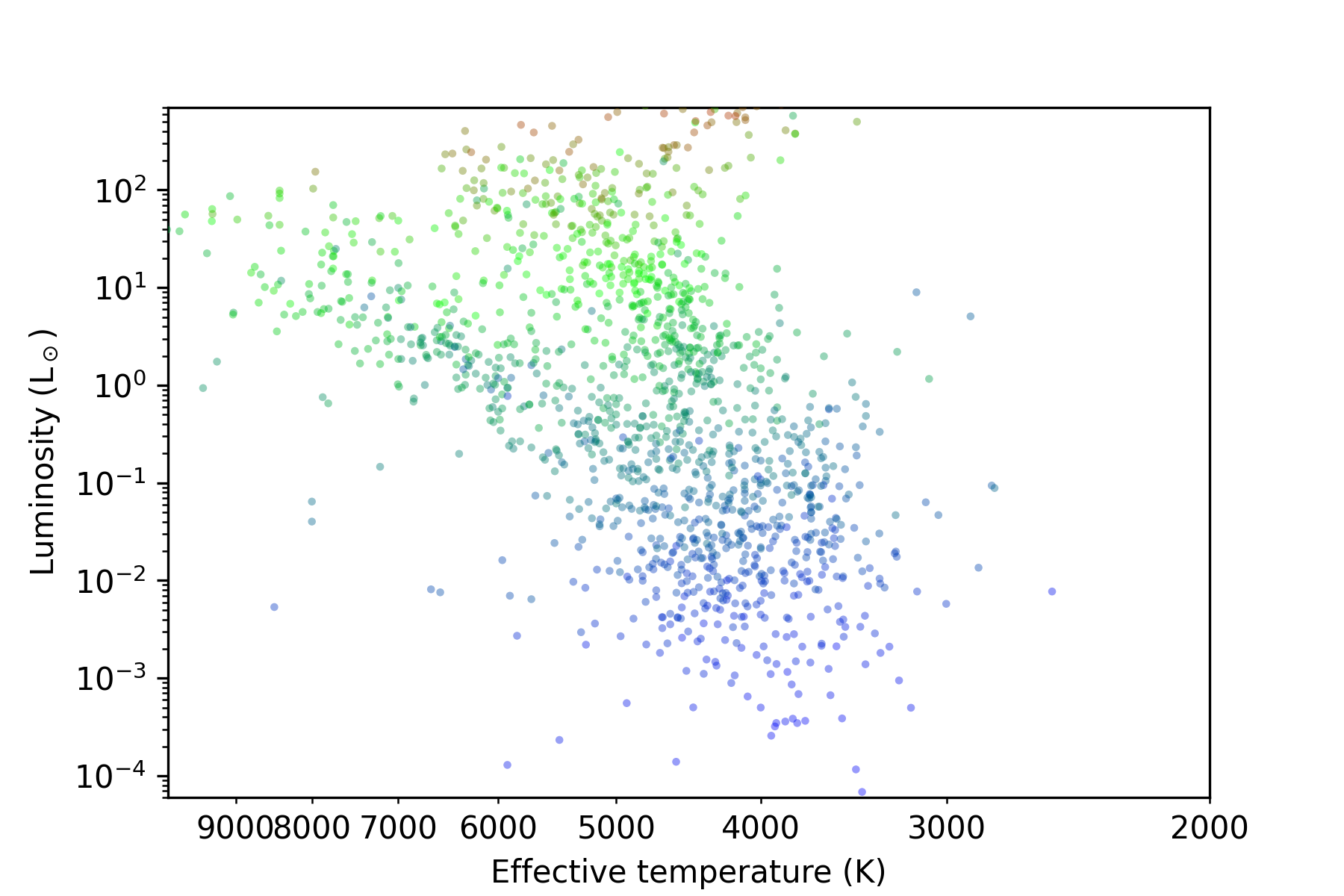}
 \caption{H--R diagrams showing (top panel) a cone of 60$^\prime$ radius centred on the North Galactic Pole, (lower panel) a cone of 3$^\prime$ radius from near the Galactic Bulge, centred on 18$^h$ --30$^\circ$. Colours, ordered from blue to red, refer to the adopted distance of the star.}
 \label{fig:HRDCrowding}
\end{figure}

Examples of these problems are shown in Figure \ref{fig:HRDCrowding}. In both the cases of the Galactic Pole and Galactic Bulge, around $\sim$10\,000 \emph{Gaia} DR3 objects fall within the search cone. However, the parallax uncertainty of those towards the Bulge are typically much greater, so fewer stars are plotted.

We have chosen a region of the Galactic Bulge with low local reddening, which is not captured by our grid model. The over-correction of reddening in this region causes the stars that do remain in the H--R diagram to be more scattered due to photometric uncertainty, and placed too warm and too bright in the H--R diagram due to over-correction of reddening.  This also causes a very poor reproduction of the giant branches, which comprise almost entirely of stars at larger distances. as the reddening from the 3D extinction cubes grows steadily in this direction with distance, while the true interstellar reddening in this narrow field appears not to. Progressive over-correction of interstellar reddening further up the giant branch ultimately leads to its slope becoming inverted.

\subsection{Variable extinction}
\label{sec:Extinction}

Regions of high extinction cause problems for extinction correction to SEDs. There are two related issues here: first, the uncertainty in the absolute extinction means that over- or under-correction of photometry can take place, causing a global offset to a region's H--R diagram; second, small-scale spatial variations in reddening are unaccounted for in extinction maps, meaning it can be difficult to properly correct for extinction on the edges of clouds.

\begin{figure}
 \includegraphics[width=\columnwidth]{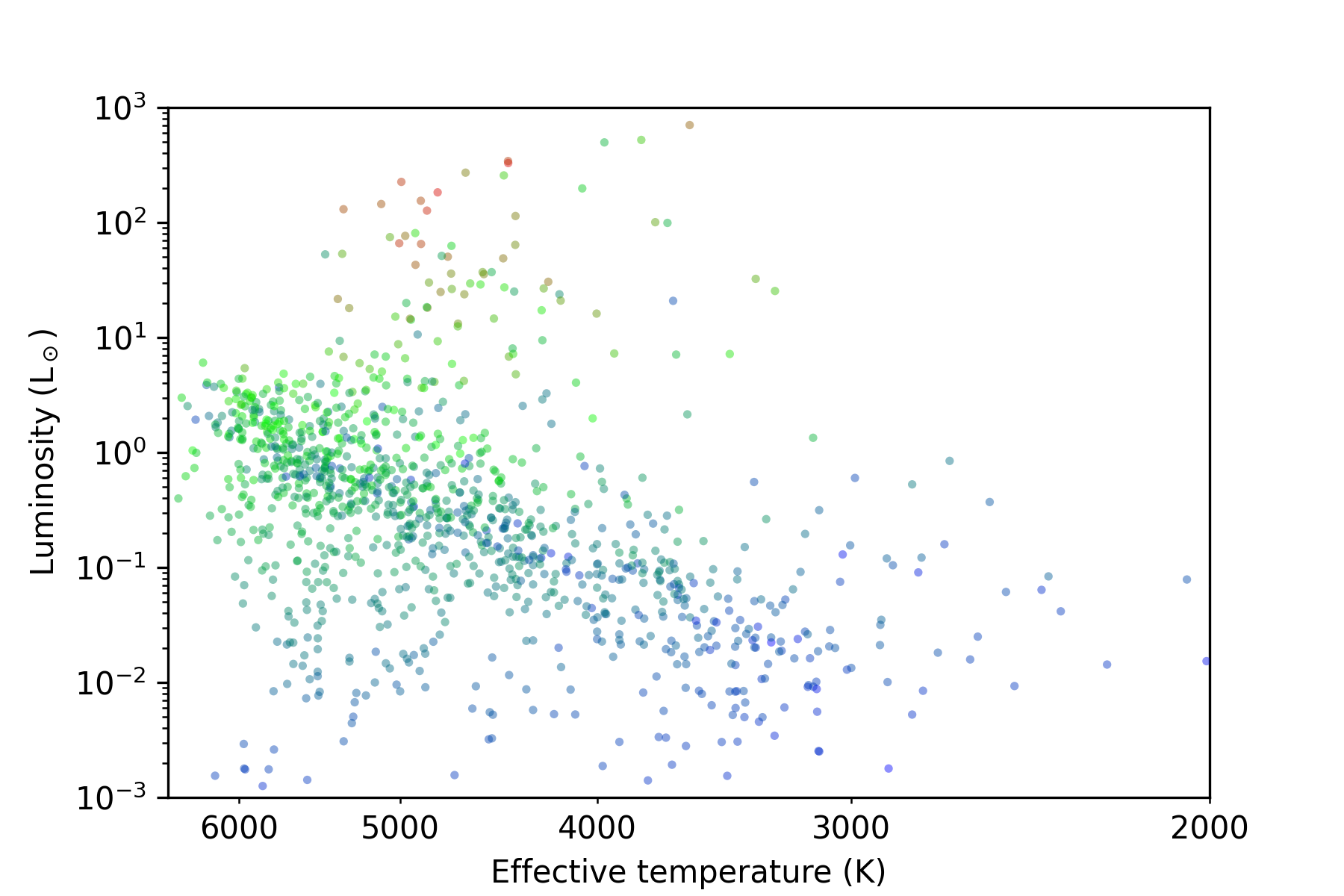}
 \caption{H--R diagram showing a $0.1 \times 1^\circ$ box transecting the Corona Australis cloud.}
 \label{fig:HRDCloud}
\end{figure}

Figure \ref{fig:HRDCloud} shows the effect of sampling stars progressively closer to a region of high extinction out of the Galactic Plane, in Corona Australis. Unlike the Galactic Bulge (Figure \ref{fig:HRDCrowding}), the CrA cloud shows a well-placed main sequence, though one which is more truncated and has higher scatter than the well-formed H--R diagram of the Galactic Pole. As stars experience increasingly higher extinction on scales finer than the extinction cube can model, the positions of the stars in the H--R diagram become shifted to cooler temperatures and slightly fainter luminosities. Since the degree of movement depends on the blackbody peak, the effect is more pronounced for hotter stars. This causes a blurring of the main sequence in this direction. Distance uncertainties exacerbate this scatter, as it places some stars at different distances within the cloud, some sources being erroneously placed in front of the major source of extinction and some behind.

\subsection{Statistical performance}
\label{sec:Performance}

\begin{figure}
 \includegraphics[width=\columnwidth]{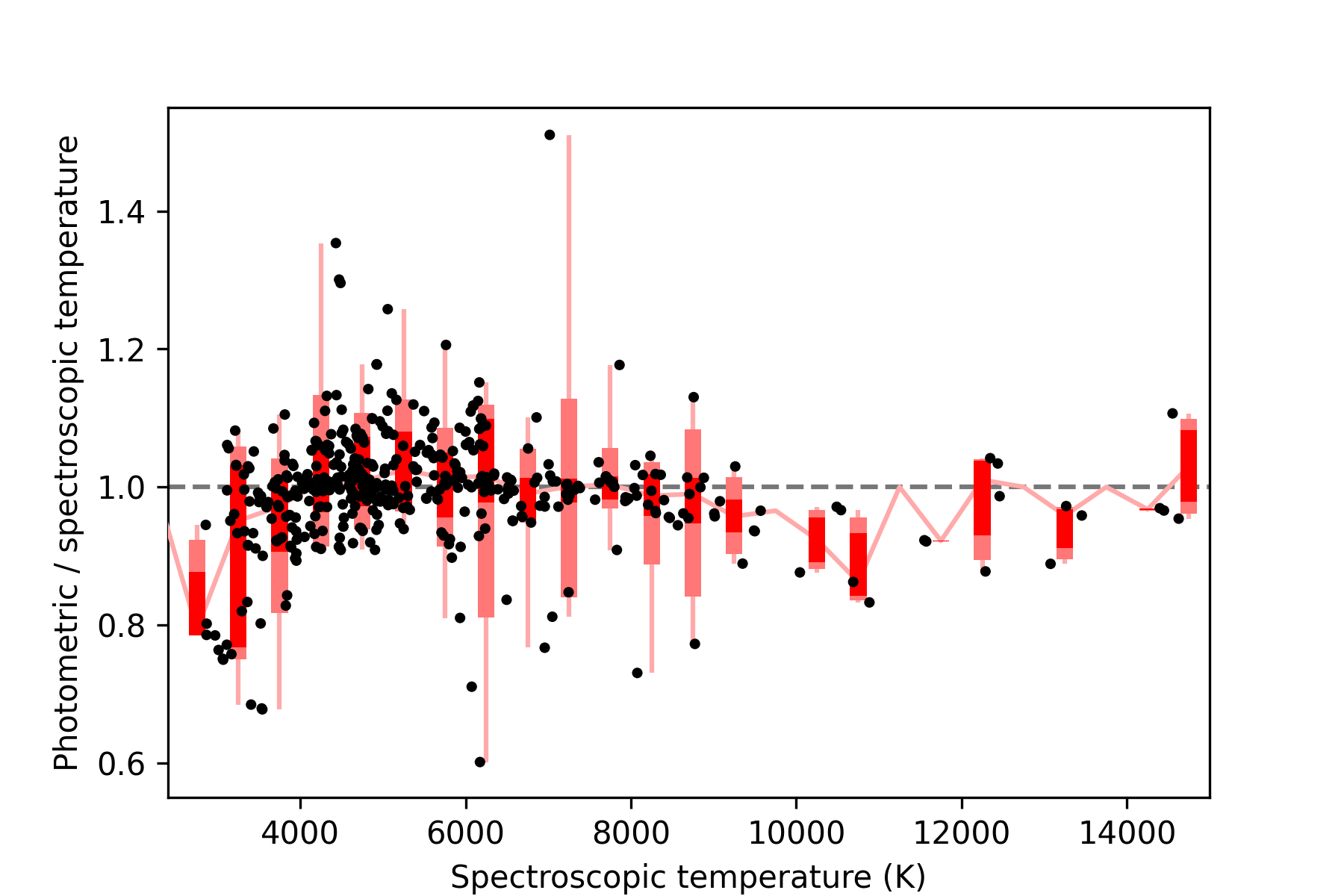}
 \caption{Ratio of fitted photometric ({\sc PySSED}) to spectroscopic (X-Shooter) temperatures for the \citet{Arentsen2019} dataset. Black dots show individual stars. Candlesticks are drawn in red, every 500 K, showing the full range (error bars), 5th to 95th centiles (light-coloured boxes) and 16th to 84th centiles (dark-coloured boxes).}
 \label{fig:XSL}
\end{figure}

To determine the overall statistical performance of {\sc PySSED}, we have analysed the set of 754 stars with stellar atmosphere characteristics from the X-Shooter Spectral Library \citep{Arentsen2019}. Of these stars, 469 had distance measurements accurate to within 50 per cent and at least three photometric datapoints to fit. Many of the remainder have poor-quality SEDs due to the problems listed above: the number of dust-enshrouded, variable stars, high-proper-motion stars, and stars in clusters or external galaxies are all above average. This therefore demonstrates how well {\sc PySSED} performs in a varied but generally quite challenging environment. More representative samples are given in the next section.

Figure \ref{fig:XSL} compares the literature spectroscopic fits to the automated fits by {\sc PySSED} in its default settings. Overall, the dataset shows a scatter around parity, with the exception of very high temperatures, where small numbers of stars make it difficult to establish a good estimate of performance, and very low temperatures, where self-extinction by dust in giant stars means that the definitions of photometric and spectroscopic temperatures diverge, as the SEDs and spectral features sample different heights in the star's atmosphere.

The distribution in Figure \ref{fig:XSL} at any given temperature is distinctly non-Gaussian, with a central group of well-fit stars, which have well-behaved SEDs and make up around half of the sample, and a series of outlying stars, which have poorly behaved SEDs and make up the other half of the sample. Most of the stars in this second set are affected by one or more of the issues highlighted above.

For example, the outlier at the top of Figure \ref{fig:XSL}, with the highest ratio, is HD\,284248: here, most of the photometry for the star is incorrect. The APASS photometry and \emph{WISE} [3.4] and [4.6] bands are correctly identified as saturated and removed. The star appears in catWISE, but the photometric data are still unsuitably corrected. The Pan-STARRS cross-match is incorrect: due to the long timespan of the Pan-STARRS survey, the correct cross-match at the actual epoch of observation is a long way from the projected position of the star at the adopted epoch of the survey, and a different cross-match is made. These points would normally be flagged and removed as outliers, except that {\sc PySSED} has (by default) been instructed to stop rejecting outliers after more than half the photometric points have been removed. It is possible that this particular fit could be improved by a magnitude-weighted cross-match or a cross match specific to the epoch of individual stars within a survey, but this is difficult without either assuming a ground truth for magnitude, or significantly adding to the computation time of the cross-matching process.

\subsection{Uncertainties and limitations}
\label{sec:Errors}

{\sc PySSED} does not currently provide uncertainties on output quantities. As seen in the previous section, the precision of the output parameters depends primarily on the accuracy of the SED fit. This accuracy is not possible to quantify robustly, as it depends on many factors. For sources with the highest uncertainties (e.g., HD\,284248, above), the accuracy of the cross-matching and correct flagging and removal of saturated and otherwise outlying points represents the dominant source of error. Sources that have (astrophysically real) SEDs that are not well represented by stellar model, such as dust-enshrouded stars and stars in nebulae, will have larger uncertainties too.

However, even for stars with well-fit SEDs, the dominant source of error is often not the photometric uncertainty on a single data point. As noted in Section \ref{sec:bbfit}, we apply an additional error to each photometric datum to account for systematic uncertainties that are not encompassed in the survey's official uncertainty budget. Such systematic errors include:
\begin{itemize}
    \item Errors resulting from stellar blending, sources belonging to data artefacts in surveys, or unflagged problems like saturation.
    \item Uncertainties in the photometric zero points of individual surveys, including any colour corrections that remain unperformed in the photometric reduction. Currently we have no means of automatically detecting whether colour corrections remain necessary to a given photometric dataset: these corrections normally affect the photometry for an individual filter by between a fraction of and a few per cent \citep[e.g.][]{Wright2010}, and the absolute calibration of magnitudes to flux is typically $\sim$2 per cent \cite[e.g.][]{Tokunaga2005}.
    \item Associated errors in filter transmission curves. This especially includes any unmeasured red or blue leaks that affect unusually cool or hot stars.
    \item Deficiencies in the stellar model atmospheres used to fit. While modern stellar model atmospheres very closely match most stars, those used here are a more generic version. {\sc PySSED} makes assumptions or estimates of stellar gravity and composition, but relies fundamentally on a grid of models and interpolation (especially linear interpolation) among this grid is imperfect. As well as minor errors in input physics, the underlying models also may not adequately represent stars with different abundances of individual elements, rotating or magnetically active stars. This adds a small and poorly quantifiable uncertainty.
    \item Uncertainties in the amount of interstellar extinction, and limitations in the resolution of interstellar extinction cubes. As highlighted in Section \ref{sec:Extinction}, extinction correction in the modern era is good, but far from perfect. Poor extinction correction, particularly in regions of strong extinction and small-scale extinction structure, can dominate the uncertainty budget. While the {\sc g-tomo} extinction cubes come with uncertainties, these uncertainties are not used currently used by {\sc PySSED} and do not account for fine structure in interstellar extinction, nor any changes in its wavelength dependence.
\end{itemize}

This litany of poorly quantified uncertainties makes constructing an error model for {\sc PySSED}'s outputs impractical, hence we have regrettably taken the decision not to provide errors for our output quantities. This also highlights the main restrictions that determine the fundamental limits to accuracy in our fits. While many (e.g., filter- or survey-specific errors) can be reduced by incorporating data from more surveys into the fit, some (e.g., reddening corrections) cannot be improved without a more fundamental treatment.


\section{Applications}
\label{sec:Applications}

\subsection{Galactic H--R diagrams}
\label{sec:Globs}

\begin{figure*}
 \includegraphics[width=\textwidth, trim=0 0cm 0 0cm, clip]{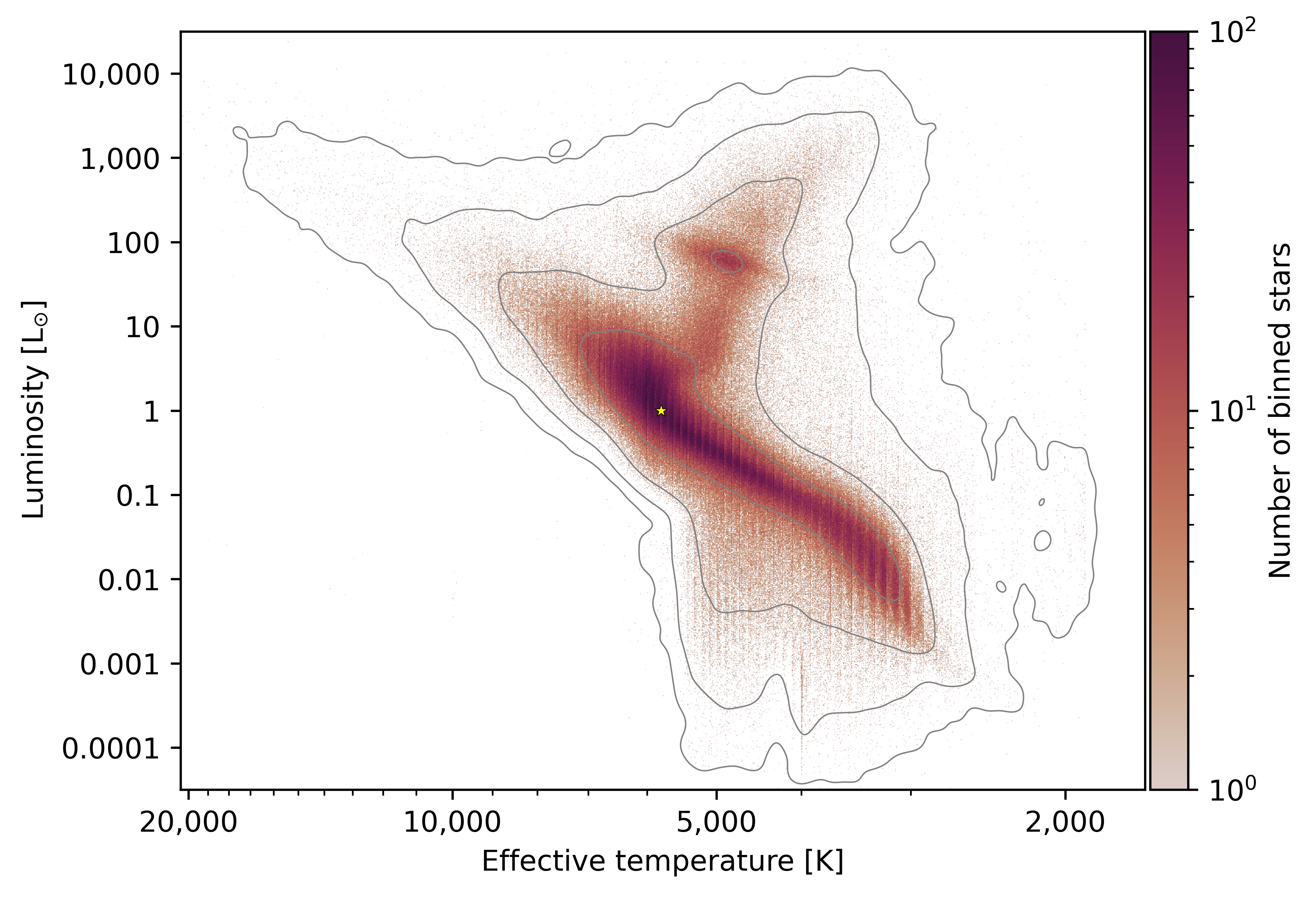}
 \caption{A density H--R diagram of 7448 random fields on the sky, with key features indicated. Isodensity contours are given for clarity and are generated by image smoothing (convolving the binned data by a 2D Gaussian kernel). The Sun is marked as a yellow dot.}
 \label{fig:HRDRandom}
\end{figure*}

As a test, we defined 7448 random positions on the sky and performed a {\sc PySSED} area search with a radius of 6$^\prime$ around each position, covering roughly 0.57 per cent of the sky. A minimum of six photometric measurements per star were required, as well as a fractional error in the \emph{Gaia} DR3 parallax of no more than more 15 per cent. Sources with parameters outside the fitted range of the {\sc bt-settl} models are not fitted: this excludes (for example) high-gravity sources, including white dwarfs.

Figure \ref{fig:HRDRandom} shows the resulting H--R diagram of the 1\,080\,531 objects with fits among 9\,949\,386 unique total \emph{Gaia} sources. The main sequence and giant branches are clearly identified. The red clump on the giant branch is clearly visible, as is the gap between the red clump and the base of the AGB. Also visible are a few cool stars (3000--5000\,K) whose luminosities lie below the main sequence. This region is dominated by white-dwarf binaries \citep[e.g.][]{Abril2020}, which may not be fit well with our single-star models. Crowding in open clusters can also contribute to this group.

Several artefacts are also present. Vertical stripes correspond to the temperatures of the model atmosphere grid, which are related to the linear interpolation performed between grid points. These are particularly common in cool objects, and arise due to inflection points across the optical range where molecular bands are strong. Noisy photometry can ``trap'' the model at these points. While a cubic or other more complex interpolation could be performed, the computational difficulty of doing this in four dimensions was judged to unacceptably slow the fitting process.

There is also some vertical spread in the main sequence, caused by parallax uncertainties, and a significant diagonal spread to the giant branch. This latter spread is partly due to a spread in metallicity, but mostly due to errors in reddening correction.

\subsection{Stellar clusters and nearby galaxies}

\subsubsection{The globular cluster NGC 104 (47 Tuc)}
\label{sec:NGC104}

\begin{figure*}
 \includegraphics[width=0.49\textwidth, trim=0.5cm 0.4cm 0.37cm 0.2cm, clip]{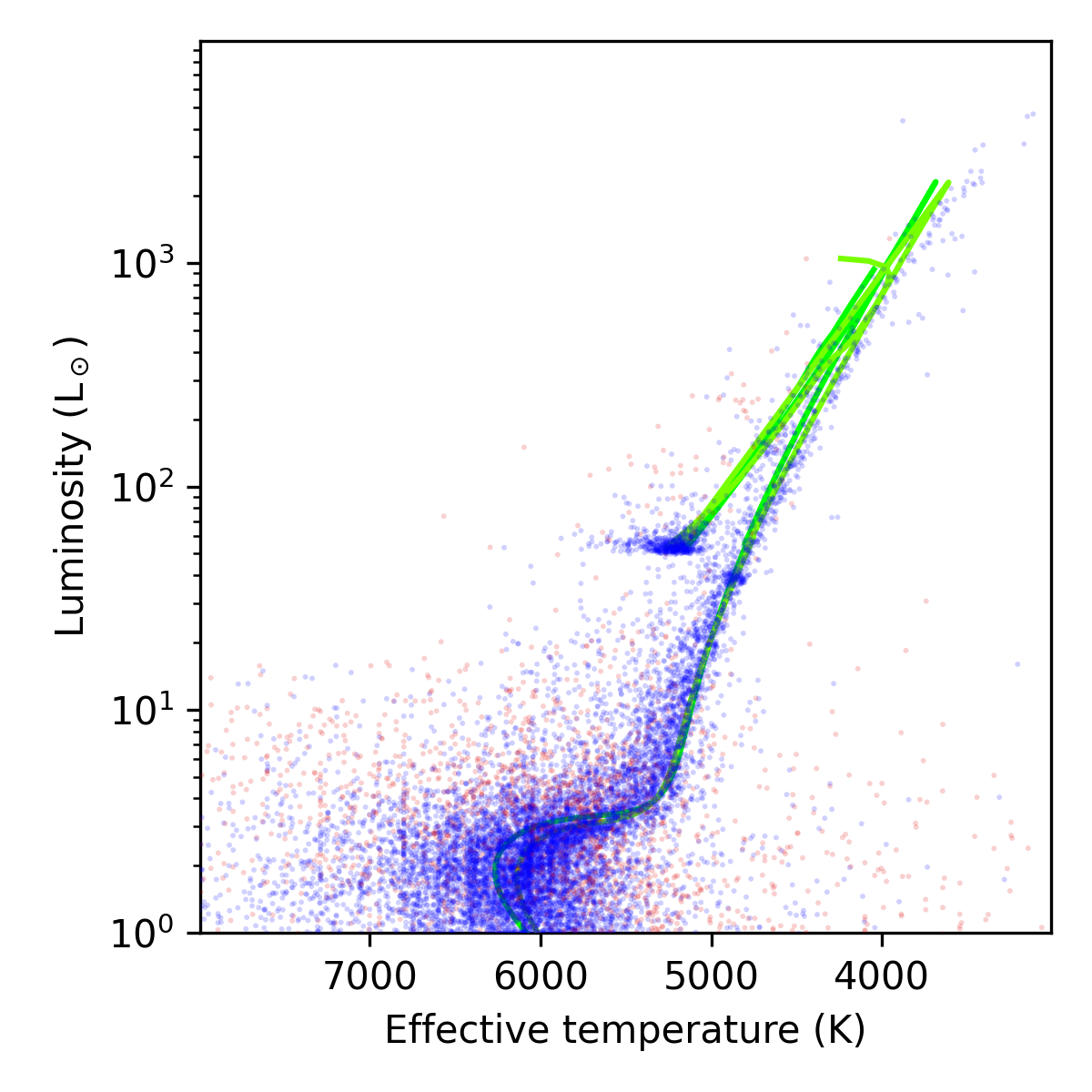}
 \includegraphics[width=0.49\textwidth, trim=0.5cm 0.4cm 0.37cm 0.2cm, clip]{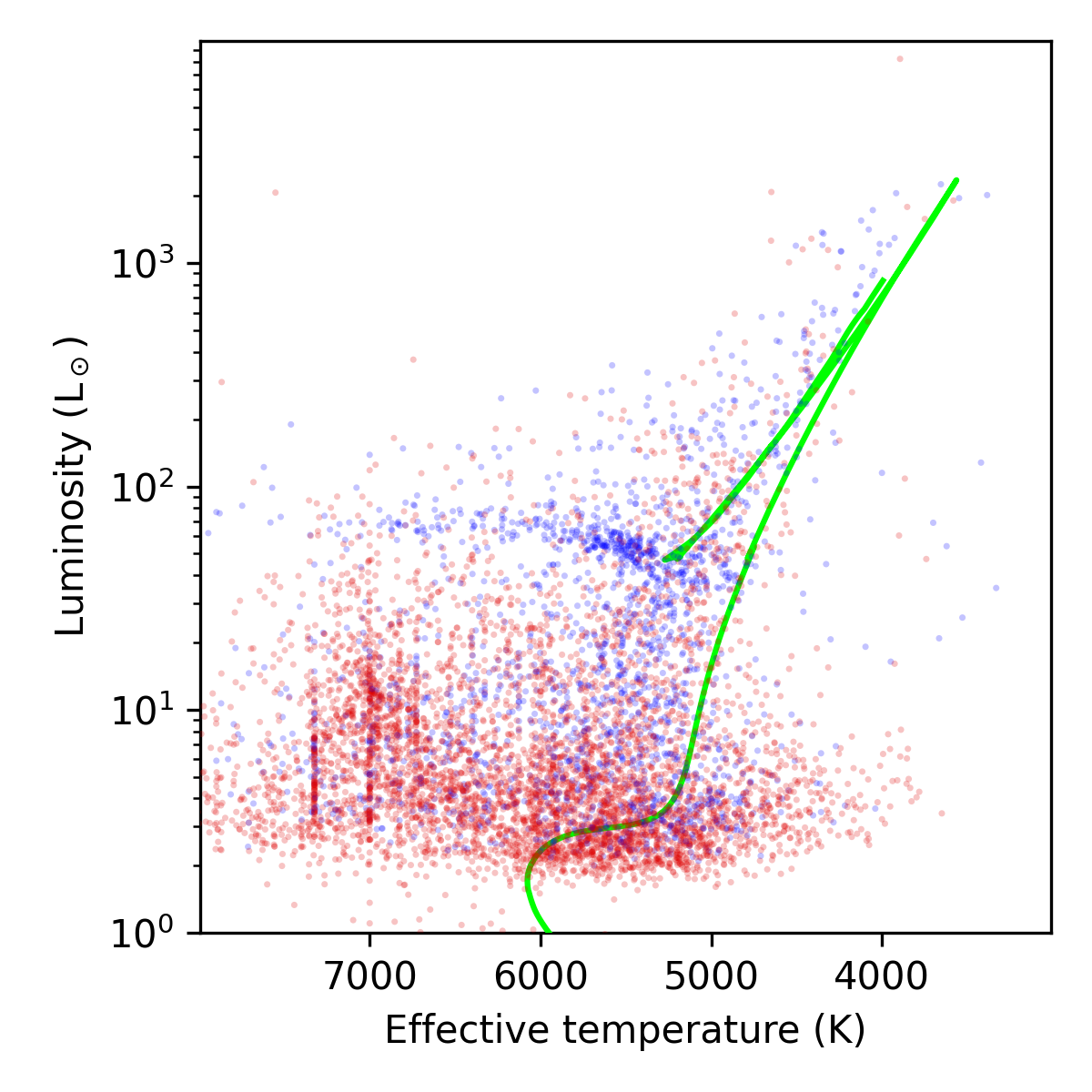}
 \caption{ H--R diagrams of the globular clusters NGC\,104 (47\,Tuc; left panel) and NGC\,6539 (right panel). Each diagram shows stars within 6$^\prime$ of the cluster centre. Stars are colour-coded by proper-motion and parallax members (blue points) and proper-motion or parallax non-members (red points). Green lines show isochrones, described in the text.}
 \label{fig:HRDglobs}
\end{figure*}

\begin{table}
    \caption{Differences between photometric and spectroscopic temperature estimates in 47 Tucanae.}
    \label{tab:47tuc}
    \centering
    \begin{tabular}{@{}l@{\ }r@{}l@{\ }r@{}l@{\ }c@{\ }l@{}}
        \hline\hline
        \multicolumn{1}{c}{Survey} &
        \multicolumn{2}{c}{Average$^1$} &
        \multicolumn{2}{c}{Median$^2$} & 
        \multicolumn{1}{c}{$N^3$} &
        \multicolumn{1}{c}{Notes}
         \\
        \multicolumn{1}{c}{\ } &
        \multicolumn{2}{c}{(K)} &
        \multicolumn{2}{c}{(K)} & 
        \multicolumn{1}{c}{\ } &
        \multicolumn{1}{c}{\ }
         \\
        \hline
    \citet{Abdurrouf2022} & $-23$&$\pm 149$ & $-18$&$^{+93}_{-106}$ & 96 & APOGEE$^4$\\
    \citet{Soubiran2016} & $174$&$\pm 576$ & $117$&$^{+217}_{-128}$ & 28 & PASTEL$^5$\\
    \citet{Hourihane2023} & $31$&$\pm 185$ & $-6$&$^{+158}_{-103}$ & 158 & \emph{Gaia}--ESO\\
    \citet{Kovalev2019} & $104$&$\pm 132$ & $57$&$^{+100}_{-33}$ & 24 & $^6$\\
    \citet{Aoki2021} & $181$&$\pm 227$ & $149$&$^{+126}_{-118}$ & 82 & $^6$\\
    \citet{Nepal2023} & $34$&$\pm 185$ & $29$&$^{+166}_{-130}$ & 96 & $^6$\\
    \citet{Carretta2009} & $112$&$\pm 128$ & $107$&$^{+43}_{-41}$ & 93 & Photometric\\
    \citet{Gratton2013} & $61$&$\pm 66$ & $50$&$^{+39}_{-43}$ & 65 & Photometric\\
    \citet{Kolomiecas2022} & $12$&$\pm 103$ & $11$&$^{+50}_{-42}$ & 122 & Photometric$^8$\\
    \citet{Mucciarelli2017} & $112$&$\pm 130$ & $106$&$^{+45}_{-40}$ & 91 & Photometric\\
    \citet{DOrazi2010} & $318$&$\pm 187$ & $318$&$^{+143}_{-98}$ & 20 & Photometric\\
    \citet{Wang2017} & $78$&$\pm 36$ & $87$&$^{+22}_{-33}$ & 49 & Photometric\\
    \citet{Dobrovolskas2014} & $222$&$\pm 193$ & $235$&$^{+151}_{-163}$ & 18 & H$\alpha^{(7)}$\\
    \citet{Lane2011} & $290$&$\pm 291$ & $248$&$^{+255}_{-140}$ & 823 & $^{9}$\\
    \citet{Cordero2014} & $-3$&$\pm 151$ & $12$&$^{+58}_{-87}$ & 52 & \vspace{2mm}\\
    {\it Total} & $169$&$\pm 266$ & $124$&$^{+245}_{-132}$ & 1827 & \\        
        \hline
        \multicolumn{7}{p{0.95\columnwidth}}{$^1$Uncertainties show standard deviation. $^2$Uncertainties show 84th and 16th centile intervals. $^3$Number of stars in common. $^4$APOGEE uses a combination of photometric and spectroscopic temperatures to produce its final estimate. $^5$PASTEL is a bibliographic sample, so contains hetrogeneous data. $^6$Uses existing data from the \emph{Gaia}--ESO survey. $^7$Only main-sequence turn-off stars, uses the H$\alpha$ wings to estimate temperature. $^8$Checked against Fe ionisation levels. $^9$Uses variant of the RAVE pipeline \citep{Kiss2007}.}\\
        \hline
    \end{tabular}
\end{table}

\begin{figure}
 \includegraphics[width=\columnwidth, trim=0 0 0 0, clip]{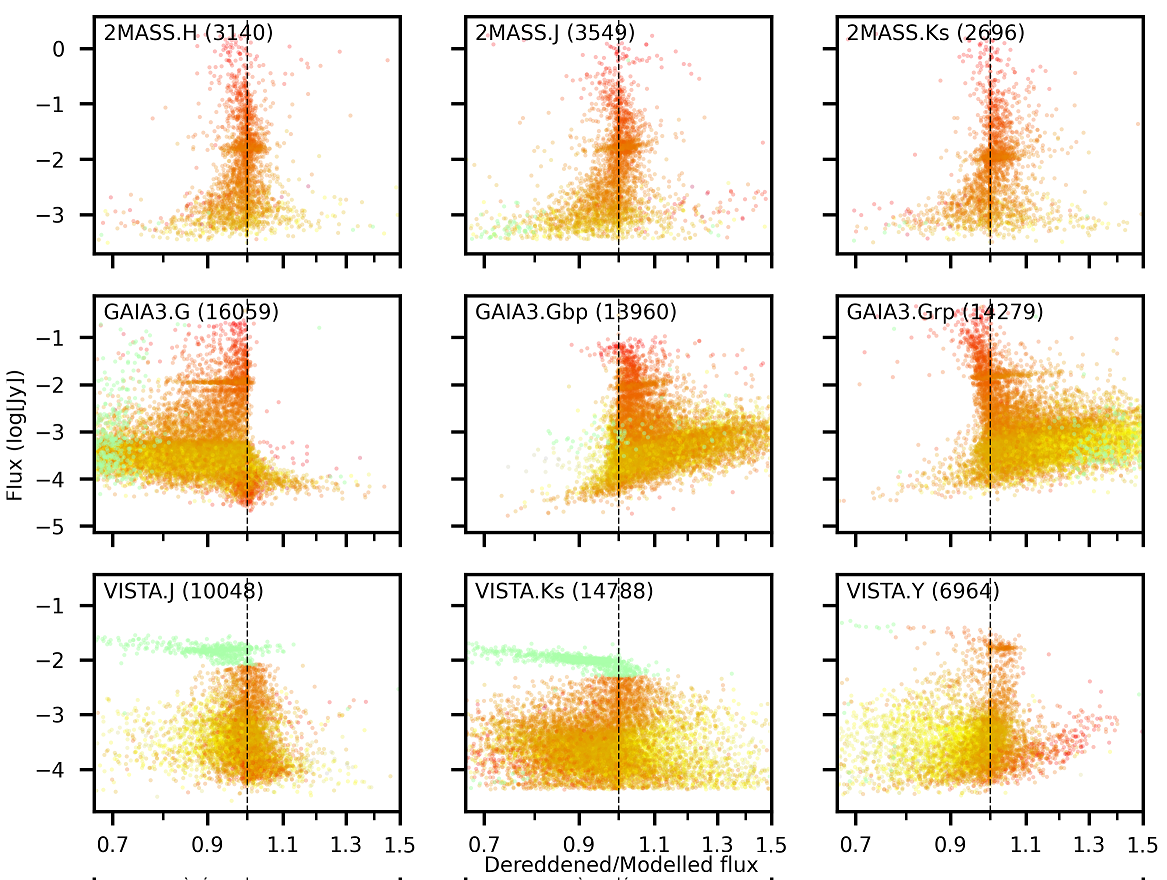}
 \includegraphics[width=\columnwidth, trim=0 0 0 0, clip]{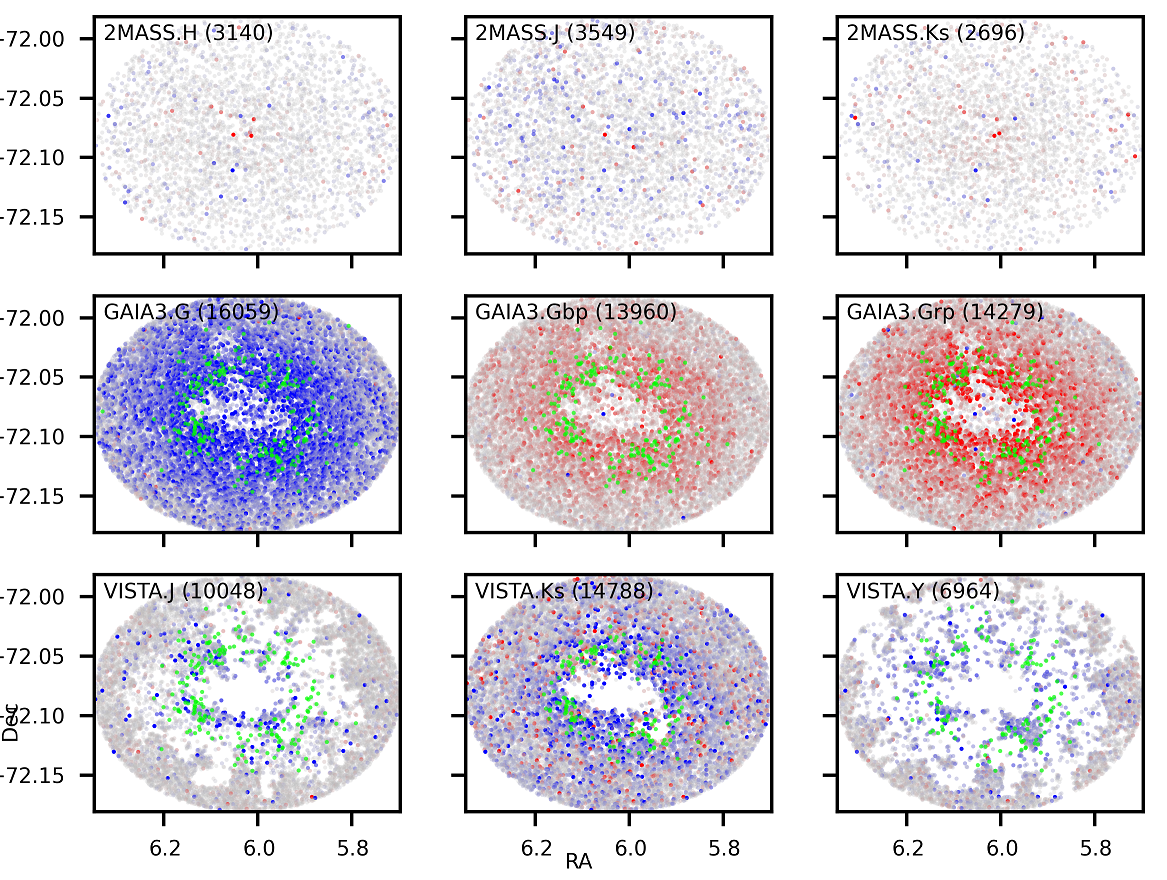}
 \includegraphics[width=\columnwidth, trim=0 0 0 0, clip]{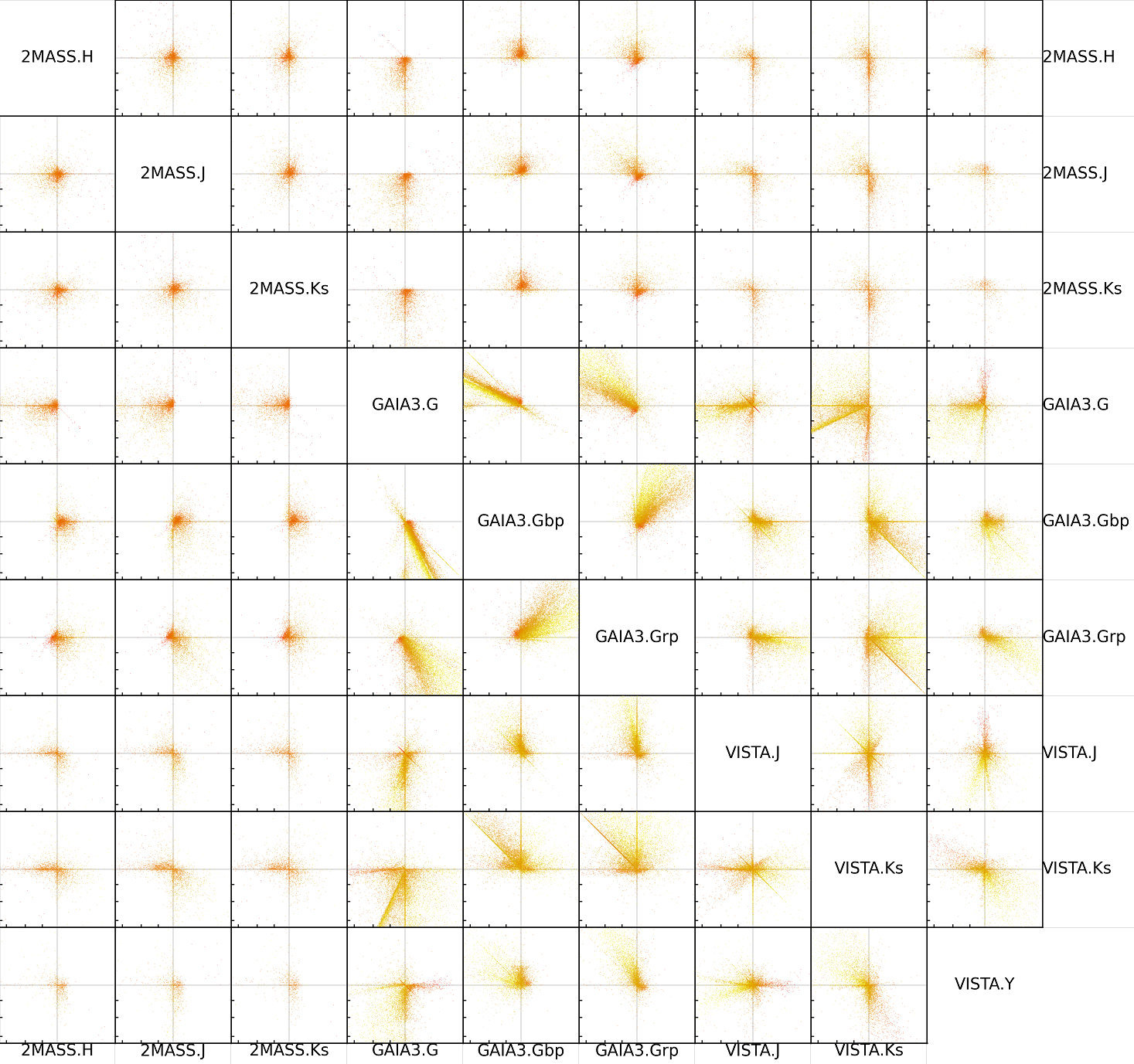}
 \caption{Excess flux diagnostic charts for NGC\,104, cropped from the original program outputs. Top panel: excess flux versus brightness; colours represent blackbody colours (red $\sim$ 3000 K, yellow $\sim$ 6000 K). Middle panel: excess flux on the sky (bluer shades show flux deficits, red shades show flux excesses). Bottom panel: excess versus excess chart for filter pairs; the axes range logarithmically from 0.67 to 1.5 in all cases, and colours are as in the top panel. Green points denote photometry discarded from the fit due to either saturation or discrepancies.}
 \label{fig:XS47}
\end{figure}

Figure \ref{fig:HRDglobs} shows the H--R diagram of two globular clusters, NGC\,104 (also known as 47 Tucanae, discussed here) and NGC\,6539 (discussed in the next section). Here, the metallicity of the cluster is fixed, and all stars with parallax uncertainties greater than 5 per cent (for NGC\,104) and 15 per cent (for NGC\,6539) are placed at the distance of the cluster \citep{Baumgardt2021}. Objects are assumed to be cluster members if they are either (1) placed in the cluster by the above uncertainty criterion, or if (2) their parallax distances are more than half that of the cluster (and are therefore not identifiably foreground stars). Stars are also only considered cluster members if their proper motions are either null or within 3\,mas\,yr$^{-1}$ of the cluster's proper motion. Only the central 6$^\prime$ of each cluster is examined, thus providing an extremely dense and challenging environment from which to extract parameters.

The H--R diagram of NGC\,104 was previously presented by \citep{McDonald2011}, which used an entirely independent set of photometry. The H--R diagram in Figure \ref{fig:HRDglobs} closely reproduces that work, but has an offset of a few tens of Kelvin to higher temperatures. This kind of offset is not unusual when performing SED-fitting work, and arises from small photometric offsets within and between photometric surveys in different regions of the sky.

While we have no ground-truth measurements against which to derive the accuracy of our fitted parameters, we can compare to both spectroscopic temperature measurements and isochrones. NGC\,104 has been well-studied spectroscopically, and spectral temperatures and other parameters have been published for over 1000 of its stars. Table \ref{tab:47tuc} identifies the deviation of our photometric temperatures from these published estimates. However, it should be noted that many of these spectral surveys do not derive their own temperature measurements, but instead estimate them from colour relations. Broadly speaking, these literature photometric studies have temperatures approximately 100 K below ours, though there is a wide scatter in their absolute temperature calibration, which is reflected in the variety of average temperature differences between the surveys.

Of the remainder, we highlight \citet{Dobrovolskas2014}, which fits main-sequence turn-off stars. The temperature here is derived from fitting H$\alpha$ wings and shows strong discrepancies from our photometrically derived values with large scatter. The scatter is partly due to the photometric confusion in the cluster core, which limits our photometric extraction. It is unclear how confusion in the spectrograph fibres affects the spectra of faint stars this close to the centre of the cluster. A large scatter is thus expected and, without other studies, we are unable to ascertain whether the 235 K offset is real. Similarly, \citet{Lane2011} provides spectra of a large number of faint giants, but the temperatures show a very large offset and significant scatter from both our values and other surveys, suggesting their parameter estimates are typically less reliable. Meanwhile, \citet{Soubiran2016} also shows a significant offset and very large scatter, which we attribute to the heterogeny of their bibliographical input data.

This leaves us with three comparison surveys: the Sloan Digital Sky Survey Apache Point Observatory Galaxy Evolution Experiment Data Release 17 (APOGEE; \citealt{Abdurrouf2022}); the \emph{Gaia}--European Southern Observatory data release version 5.1 (\emph{Gaia}--ESO; \citealt{Hourihane2023}); and \citet{Cordero2014}. To this, we also add the photometric calibration of \citet{Kolomiecas2022} which, like \citet{Cordero2014}, made use of the Fe\,{\sc ii}/Fe\,{\sc i} ratio as a check on their derivation of $T_{\rm eff}$. These four studies consistently show agreement with our photometric temperatures, with median deviations of no more than $\pm$18\,K, and scatter of $\sim$100\,K or less. Since \citet{Cordero2014} used long, high-resolution spectra and solely used iron-abundance ratios to derive stellar temperature, we find it the most convincing and independent comparison: this therefore sets our expected minimum absolute uncertainty in the photometric temperature derived from {\sc PySSED}, of (after averaging errors) $\sim\pm$70 K for stars with good photometry, where distance, mass and extinction are known. For comparison, the temperature scatter of the red clump is $\sim\pm$60 K, some of which is astrophysical due to multiple populations within the cluster \citep[e.g][]{Johnson2015}, which sets an upper limit to the minimum internal uncertainty achievable with {\sc PySSED}.

Figure \ref{fig:HRDglobs} also includes isochrones from the ``Bag of Stellar Tracks and Isochrones'' (BaSTI; \citealt{Hidalgo2018}) and Padova \citep{Bressan2012} codes. These models are computed for [Fe/H] = --0.72 dex \citep{Harris2010}, similar to the mean metallicities of [Fe/H] = --0.76 to --0.73 dex found by the four studies mentioned in the previous paragraph. These isochrones are solar-scaled: while NGC\,104 has an $\alpha$-element enhancement, it is comparatively mild ([$\alpha$/Fe] $\approx$ 0.19 dex; \citealt{Abdurrouf2022}). In general, the isochrones match with the data well, though it should be noted that correctly accounting for the $\alpha$-enhancement would move the isochrones some tens of Kelvin cooler. Treatment of the upper giant branch depends on the exact mass-loss, diffusion and overshooting treatments applied (which we do not attempt to reproduce here), and on the dust production by the cluster's AGB stars \citep[e.g.][]{McDonald2011c}. Similarly, behaviour of the main-sequence turn-off also depends on the cluster's age, which we have simply assumed to be 11.5 Gyr. Overall, the isochrone therefore reproduces the H--R diagram of NGC\,104 well, given its expected shortcomings and the fact that we have not attempted to fit any parameters.

Despite this good match, the photometry suffers significantly from the crowding in the cluster centre. This can be seen in the diagnostic plots, portions of which are reproduced in Figure \ref{fig:XS47}. Identifying the cause of discrepancies from the model is not always trivial, as each filter affects the fit for the others.

In this case, the top panel of Figure \ref{fig:XS47} shows that we achieve a good fit for brighter stars, once the saturation in VISTA $J$ and $K_{\rm s}$ has been masked. There is also generally a good agreement for the 2MASS photometry down to $\sim$1\,mJy, with the model matching the observed flux to within a few percent. However, particularly for faint stars, there is a tension between \emph{Gaia} $G$ and VISTA $Y$ and $K_{\rm s}$, where the model over-predicts the flux for fainter stars, and \emph{Gaia} $B_{\rm p}$ and $R_{\rm p}$, where the model under-predicts the flux. This is not a case of limiting flux, as the faintest stars in \emph{Gaia} $G$ are correctly modelled. It does, however, have a temperature dependence, with the yellower stars being more significantly dispersed either side of the parity line.

The middle panel of Figure \ref{fig:XS47} shows that this is strongly related to crowding: the redder and bluer colours intensify towards the cluster centre, where many surveys give up on photometric extraction, but remain a fairly neutral colour on the outside of the sampled cone. The bottom panel of Figure \ref{fig:XS47} shows this scatter to be highly directional in the individual excess--excess plots, indicating that this is not photometric noise, but a systematic problem with the input data. This suggests that one or more of the \emph{Gaia} or VISTA bands suffers from poor photometry of faint stars in the cluster core due to the heavy blending. Removing each of the six bands sequentially and running a test sample could identify the problematic band and allow a fix (e.g., a magnitude cut) to be implemented if better precision was required, and additional, deeper photometry could be added to better sample the main sequence.

\subsubsection{The globular cluster NGC 6539}
\label{sec:NGC6539}

The right-hand panel of Figure \ref{fig:HRDglobs} shows the cluster NGC\,6539, a smaller cluster than NGC\,104, which lies in a dense field of foreground stars and has an extinction of $E(B-V)=1.02$ mag \citep{Harris2010}. Consequently, the giant branch is more poorly populated and is harder to discern from the foreground stars (red points) and, despite the cluster's being only 1.5 times further away, the H--R diagram does not extend to such deep magnitudes.

The most pertinent difference in the H--R diagram, however, is that the measured stars lie to the hotter side of the (solar scaled) BaSTI isochrone. The diagnostic excess flux diagrams for this cluster are cleaner than NGC\,104, indicating that the quality of the photometric fitting is better than NGC\,104. The hotter temperatures of our fits therefore lie in the assumptions we have made about the cluster's properties in either the fit or the isochrone.

While the properties of NGC\,6539 are not as well measured as those of NGC\,104, its metallicity is known, being close to [Fe/H] = --0.63 dex \citep{Harris2010}, i.e., slightly more metal-rich than NGC\,104. Being a metal-rich cluster, we do not expect an extended horizontal branch, and there is no extension in colour--magnitude diagrams \citep{Vasiliev2021}, yet one is seen in our H--R diagram. Despite the apparently good fits, the giant branch exhibits much more scatter than seen in NGC\,104. We can explain this by inferring an incorrect estimate of the extinction towards the cluster, either in the total visual extinction ($A_V$) or its spectral variation ($R_V$). Either could affect any metallicity estimates derived from the cluster's colour--magnitude diagrams. To explain the scatter, we could also invoke a small differential extinction across the cluster, which will be on too fine a scale to be mapped by our 3D extinction grid. NGC\,6539 therefore represents an example of how this analysis can improve the quality of parameter estimates towards globular clusters, by simultaneous fitting of prior parameter distributions in both data (e.g., extinction) and isochrones (e.g., metallicity).

\subsubsection{The Sagittarius dwarf spheroidal (Sgr dSph) galaxy}

\begin{figure}
  \centering
    \includegraphics[width=\columnwidth]{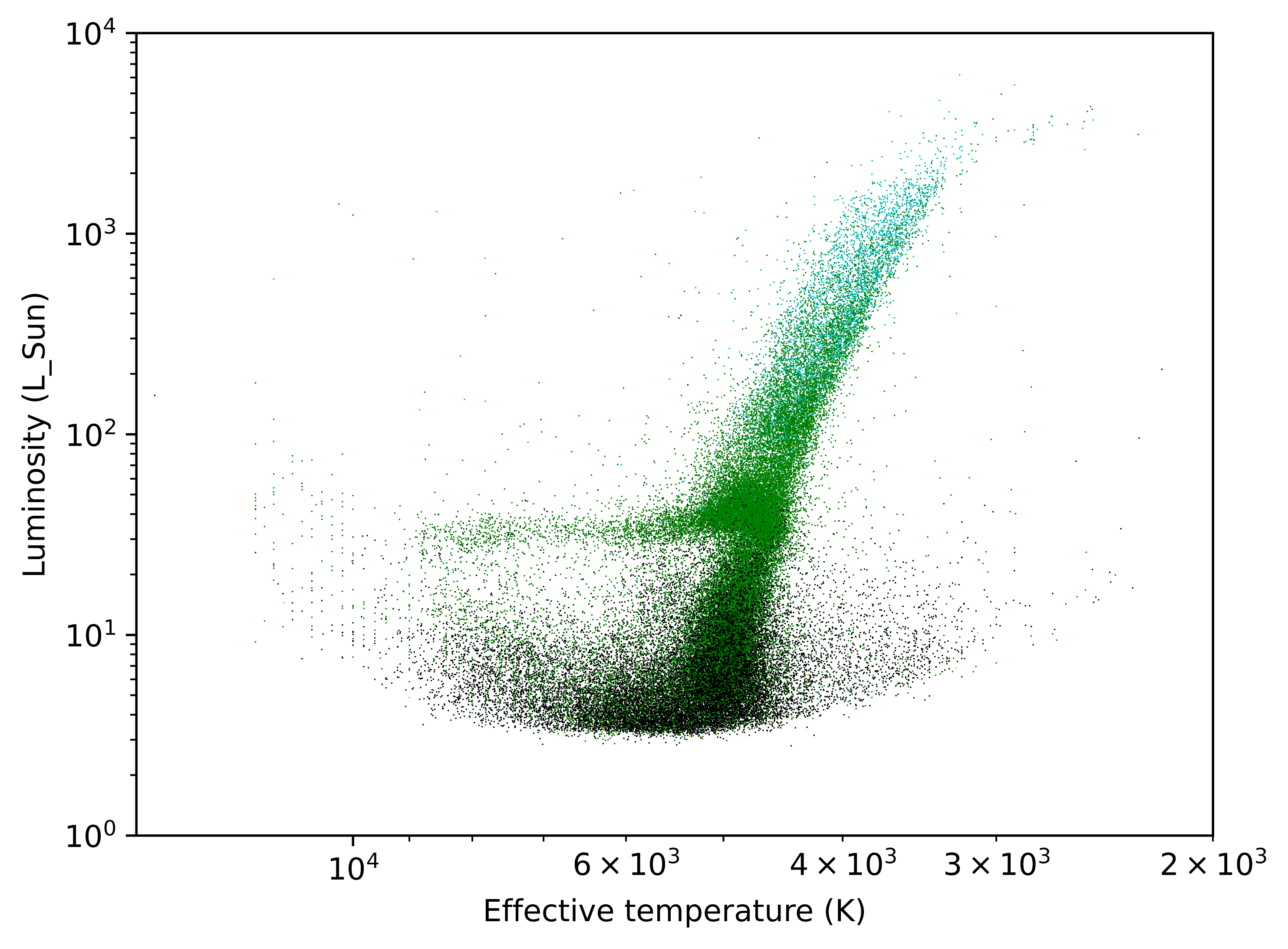}
    \includegraphics[width=\columnwidth]{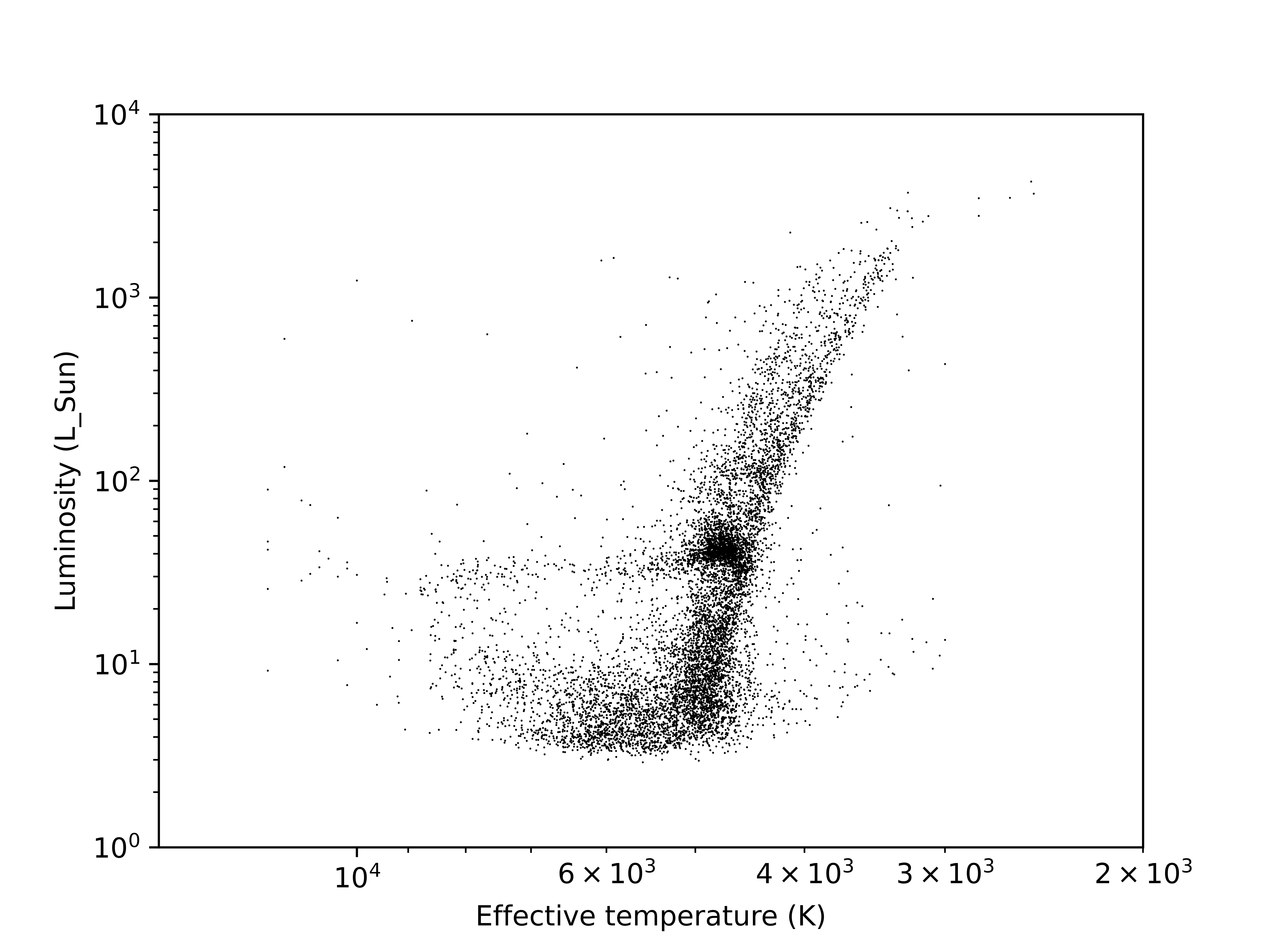}
\caption{H--R diagrams of fields covering the Sgr dSph galaxy. The top panel shows a composite of the H--R diagrams of cone searches spread across the main body of the galaxy. All objects are assumed to be at 24\,kpc distance, but are colour-coded by their best-estimate distance in \citep{BailerJones2021} (green representing 5--10\,kpc, cyan 10--24\,kpc, and black for objects without parallax information. The bottom panel shows the H--R diagram of one of these fields, centred on the globular cluster M54, which lies at the centre of the galaxy.}
\label{fig:SgrdSph}
\end{figure}

Resolved-star populations in external galaxies are continuously refining our understanding of stellar evolution in different environments \citep[e.g.][]{Boyer2017}. Parameter determinations using multi-wavelength data from sources such as the \emph{Euclid}, \emph{Roman} and \emph{James Webb Space Telescopes} will form a crucial part of this refinement.

The Sgr dSph galaxy is the nearest\footnote{The Canis Minor overdensity is closer, but its nature as a galaxy is debated \citep[e.g.][]{LopezCorredoira2012}.} satellite galaxy of the Milky Way, but was only discovered recently \citep{Ibata1995} due to strong contamination from the Galactic Bulge \citep[e.g.][]{McDonald2013}. At these distances, \emph{Gaia} parallaxes are insufficiently precise to separate these two populations, particularly for faint stars and in more crowded regions.

A series of 22 {\sc PySSED} cone searches across the galaxy were conducted from which candidate members were selected \emph{a posteriori}. Of these, 19 fields were spread across the face of the galaxy, and two larger fields were chosen towards the galaxy's tail, at higher Galactic latitudes. The fields cover a total of 30.4 square degrees on the sky. Candidate members were selected based on having \emph{Gaia} proper motions of $-3.270 < \mu_\alpha < -2.202$ and $-2.179 < \mu_\delta < -0.791$\,mas\,yr$^{-1}$. This list was then refined to exclude objects with measurable parallaxes ($\varpi$), after parallax uncertainties ($\epsilon_\varpi$) are accounted for, by retaining only stars with $\varpi + \epsilon_\varpi < 0.1$\,mas, or where the fractional error in parallax distance (direct from \emph{Gaia} DR3 or via \citet{BailerJones2021} as appropriate) exceeds 50 per cent, or where no recorded parallax information is available. Finally, faint \emph{Gaia} sources with $G > 20.5$ mag were filtered out. Remaining objects were then set to a distance of 24\,kpc and treated as members of the Sgr dSph. While these filters (particularly the proper-motion filter) are expected to be reasonably successful at removing contamination from Galactic disc and bulge stars, they will not be completely effective. No other adjustments were made to the default inputs to {\sc PySSED} for this demonstration. This includes the assumed metallicity: the Sgr dSph is metal-poor, but it has a range of metallicities \citep[e.g.][]{Sills2019}. 

The resulting H--R diagram is shown in the top panel of Figure \ref{fig:SgrdSph}. As explained below, the colour gradient seen in this figure reflects the accuracy of the parallax information and its transference to the distance scale of \citet{BailerJones2021}: this highlights the issues with relying entirely on an incomplete Galactic model, particularly in regions of the sky containing objects outside the Galaxy,and required setting the distance to 24\,kpc. Working from the bottom of the diagram upwards, the photometric limit of \emph{Gaia} and lack of data from other surveys causes a significant spread at the bottom of the H--R diagram, where stars are coloured black due to lack of reliable parallax data. The main-sequence turnoff and sub-giant branch lie on the limit of detection. The horizontal branch can be clearly seen, extending left to around 9000\,K. This range includes a number of known RR Lyrae variables \citep{McDonald2014,Ferguson2020}, which should fill the gap between effective temperatures of $T_{\rm eff} \approx 5600 - 7100$\,K. While the RGB sequence below the horizontal branch is relatively narrow, the sequence above the giant branch is wider, in part due to the addition of the AGB component. Stars in this regime generally have parallaxes, but they have large uncertainties, and are therefore placed by \citet{BailerJones2021} at closer distances due to the Galactic model used in this paper. The upper giant branches are clearly defined, with distances in \citet{BailerJones2021} becoming more distant as parallaxes become better defined. A clear RGB tip is seen just above 2000\,L$_\odot$. A faint sequence of AGB stars can be seen above the RGB tip, veering to cooler temperatures. This region contains a number of increasingly obscured carbon stars \citep[e.g.][]{McDonald2012b}, hence this apparent cooling is exacerbated by the appearance of dust around these stars, which raises the $\tau=1$ opacity layer in their atmospheres, and decreases their photometric temperature derived from SEDs below their spectroscopically derived temperatures.

The bottom panel of Figure \ref{fig:SgrdSph} shows the same diagram, restricted to the 30$^\prime$ cone around the globular cluster M54, which is taken to be the centre of the Sgr dSph. This cluster is substantially more metal-poor than the bulk of the Sgr dSph \citep[e.g.][]{Siegel2007}. This leads to the generation of two sequences in the H--R diagram, which can be seen close to each other at the base of the RGB, but diverge towards the RGB tip.

\subsection{Extinction fitting}
\label{sec:ExtFitting}

\begin{figure}
    \begin{center}
    \includegraphics[width=\columnwidth]{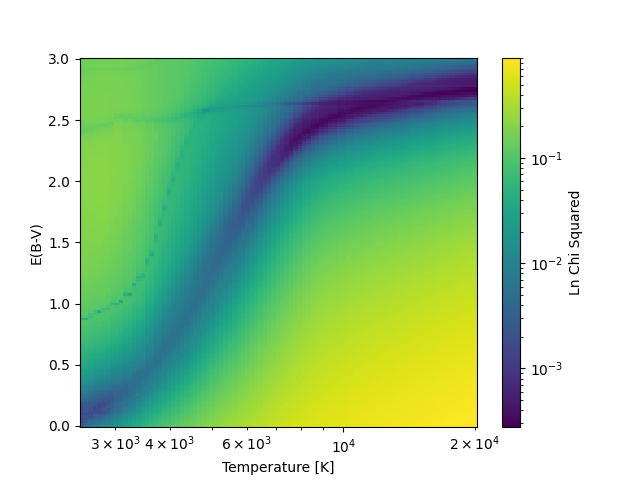}
    \includegraphics[width=\columnwidth]{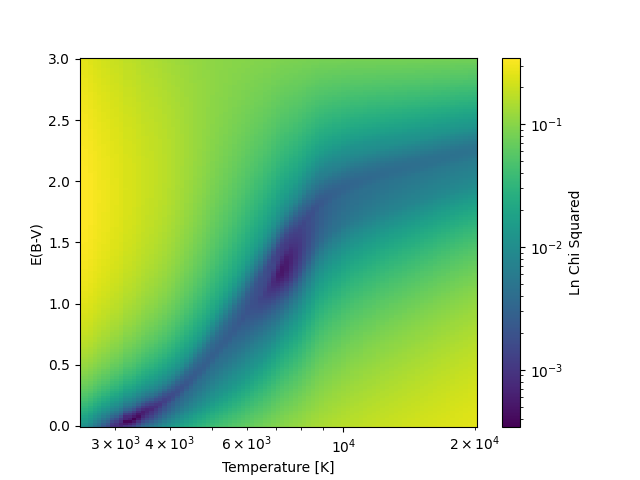}
    \caption{Goodness-of-fit maps for a joint temperature--extinction fit to \emph{Gaia} DR3 5973805862377353472 (top panel) and 5973805862391891712 (bottom panel). The goodness of fit is assessed by a modified $\chi^2$ which is weighted towards wavelengths close to the peak of the SED. A logarithmic colour scale is used. The vertical striping in the top panel is an artefact of the outlier rejection process.}
    \label{fig:extchisq}
    \end{center}
\end{figure}

\begin{figure}
    \begin{center}
    \includegraphics[width=\columnwidth]{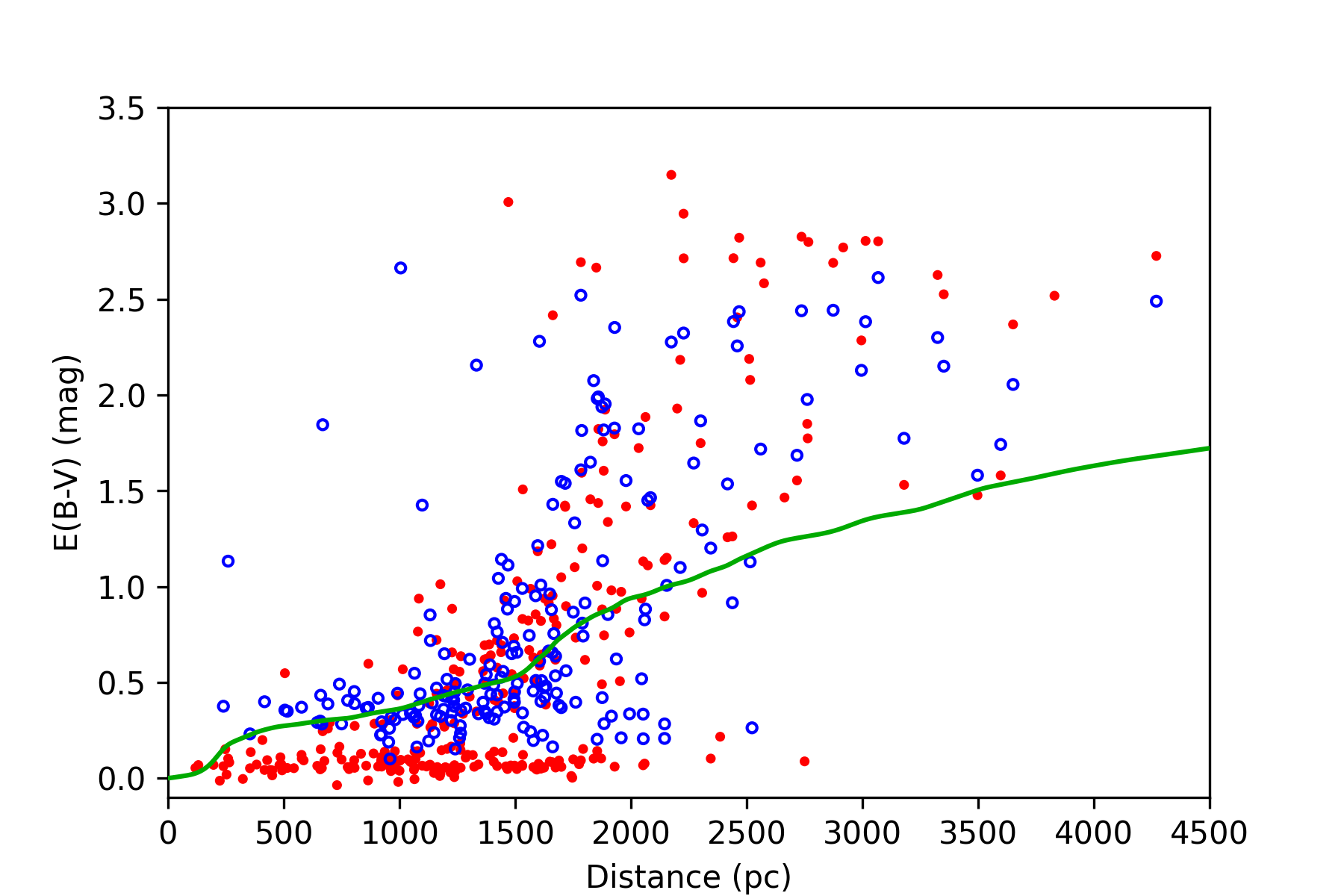}
    \caption{Fitted $E(B-V)$ versus distance for a field near to the planetary nebula NGC\,6302, centred of $\alpha = 258.425409$ deg, $\delta = -37.036262$ deg, with a radius of 196$^{\prime\prime}$. Red, filled points show the $E(B-V)$ derived from {\sc PySSED}, while blue, hollow points show $E(B_{\rm P}-R_{\rm P})$ from \emph{Gaia} DR3's GSP-Phot {\sc aeneas} pipeline, divided by a factor of 1.46 to account for the difference in extinction between the filters. The green line shows the corresponding assumption from the {\sc G-Tomo} extinction cube \citep{Lallement2022}.}
    \label{fig:NGC6302}
    \end{center}
\end{figure}

The fitted temperature and dereddening of a star are closely correlated (e.g., Figure \ref{fig:extchisq}). For this reason, and to speed computation, correcting for extinction in {\sc PySSED} is performed using a fixed (distance dependent) value derived from 3D dust maps, where the aggregated estimates of extinction from multiple stars are estimated. However, these have limited resolution and and cannot capture the effects of smaller-scale dust distribution (see Sections \ref{sec:Extinction} and \ref{sec:NGC6539}). In these cases, a better solution is to jointly fit the temperature and extinction.

If a prior in stellar temperature can be fixed using (e.g.) spectroscopic data, then the extinction is relatively easy to derive. However, in a blind fit, it is difficult to accurately determine extinction for cool stars, while for hot stars the temperature becomes entirely degenerate.

This is illustrated in Figure \ref{fig:extchisq}, which shows the blind temperature--extinction fits to two sources, Gaia DR3 5973805862377353472 and 5973805862391891712, which have respectively been fitted with high and low extinction. The first source has a well-fitted extinction of $E(B-V) \sim 2.8$ mag, but is sufficiently hot that the temperature is unconstrained. The second source has a defined temperature, but shows a degeneracy between two solutions (one with $T_{\rm eff} \sim 3200$ K and $E(B-V) \sim 0.0$ mag, and one with $T_{\rm eff} \sim 7000$ K and $E(B-V) \sim 1.4$ mag). These cases show that simultaneous fitting of temperature and extinction are at least partly possible with SED fitting in this manner, but there remains a strong correlation between the two. In cases where temperature can be defined (e.g., from \emph{Gaia} $B_{\rm P}$ and $R_{\rm P}$ spectra), more precise extinction measures can be estimated.

The two stars in Figure \ref{fig:extchisq} are both drawn from a field near the planetary nebula NGC\,6302. Figure \ref{fig:NGC6302} shows a comparison between $E(B-V)$ as fitted from \emph{Gaia} DR3 $B_{\rm P}/R_{\rm P}$ spectrophotometry, and from {\sc PySSED} joint temperature--extinction fitting for these and other stars in the region. The corresponding extinction profile from the {\sc G-Tomo} 3D extinction cubes used by {\sc PySSED} is also shown. Previously published extinction estimates towards the nebula have estimated $E(B-V) = 0.34$ mag at $\sim$700 pc and $dE(B-V)/ds = 0.34$\,mag\,kpc$^{-1}$ \citep{Matsuura2005}.

Overall, the \emph{Gaia}, {\sc PySSED} and {\sc G-Tomo} reddening estimates all reproduce two screens: one within 500 pc with a density of $\sim$0.3 mag, and one at around 1500 pc, with a much higher density. However, optical images of the $\sim 3^\prime$ field show considerable differential extinction across it. Based on the uniformity of bright blue stars in the image, this differential extinction appears to stem from distant structures (at or beyond 1500 pc).

Looking in more detail, the {\sc G-Tomo} cube exhibits a lower reddening beyond 1500 pc, while the other two have better agreement. At distance below 1500 pc, there are spuriously high extinctions in the {\sc Gaia} GSP-Phot fit that {\sc PySSED} correctly reproduces at low extinction. However, {\sc PySSED} suffers from a degeneracy in some stars, whereby very low extinctions can be erroneously fit for moderate-extinction sources.

While joint extinction fitting holds promise, it is not robust enough to use in most settings, thus is not made an ``official'' feature of this version of {\sc PySSED}. Nevertheless the code to perform this fitting is released with {\sc PySSED} for completeness. A possible future solution could be using a 3D extinction cube to develop an informed prior when fitting individual stars.

\subsection{Narrow-band spectro-photometry}
\label{sec: Narrow-band spectro-photometry}

The power of SED fitting increases with the quality of the photometry used. A major improvement in this field is expected to come from the J-PAS (Javalambre Physics of the Accelerating Universe Astrophysical Survey)\footnote{\url{https://www.j-pas.org/survey}}. This survey is covering 8000 square degrees of sky with 56 narrow-band filters across the optical spectrum. The miniJPAS survey \citep{benitez2014jpas} represents an early data release from this project. Data from this survey was sourced from the miniJPAS data access services\footnote{\url{https://archive.cefca.es/catalogues/minijpas-pdr20191}}, and the filter profiles added into PySSED from the SVO FPS.

\begin{figure}
      \centering
      \includegraphics[width=\columnwidth]{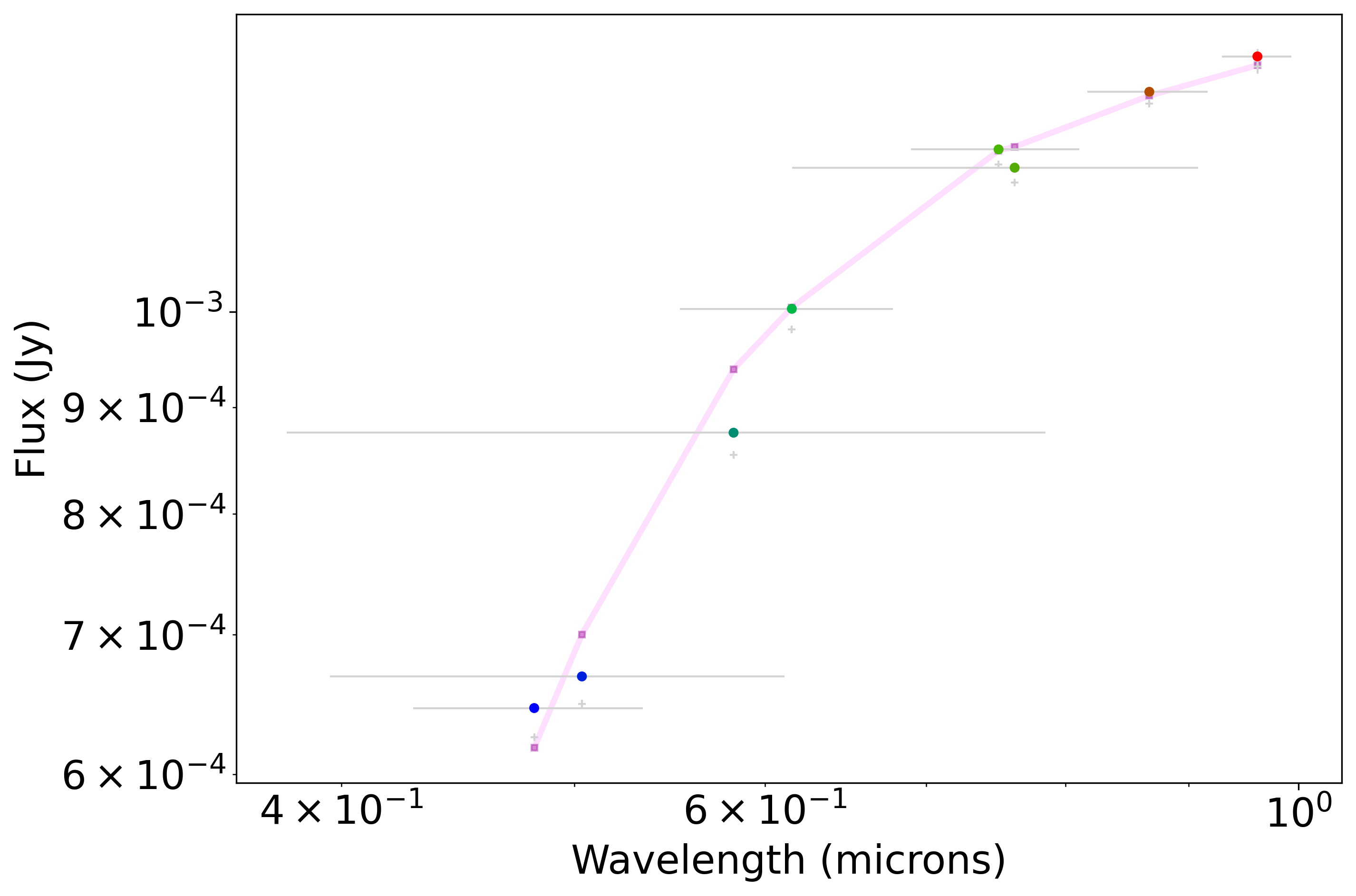}
      \includegraphics[width=\columnwidth]{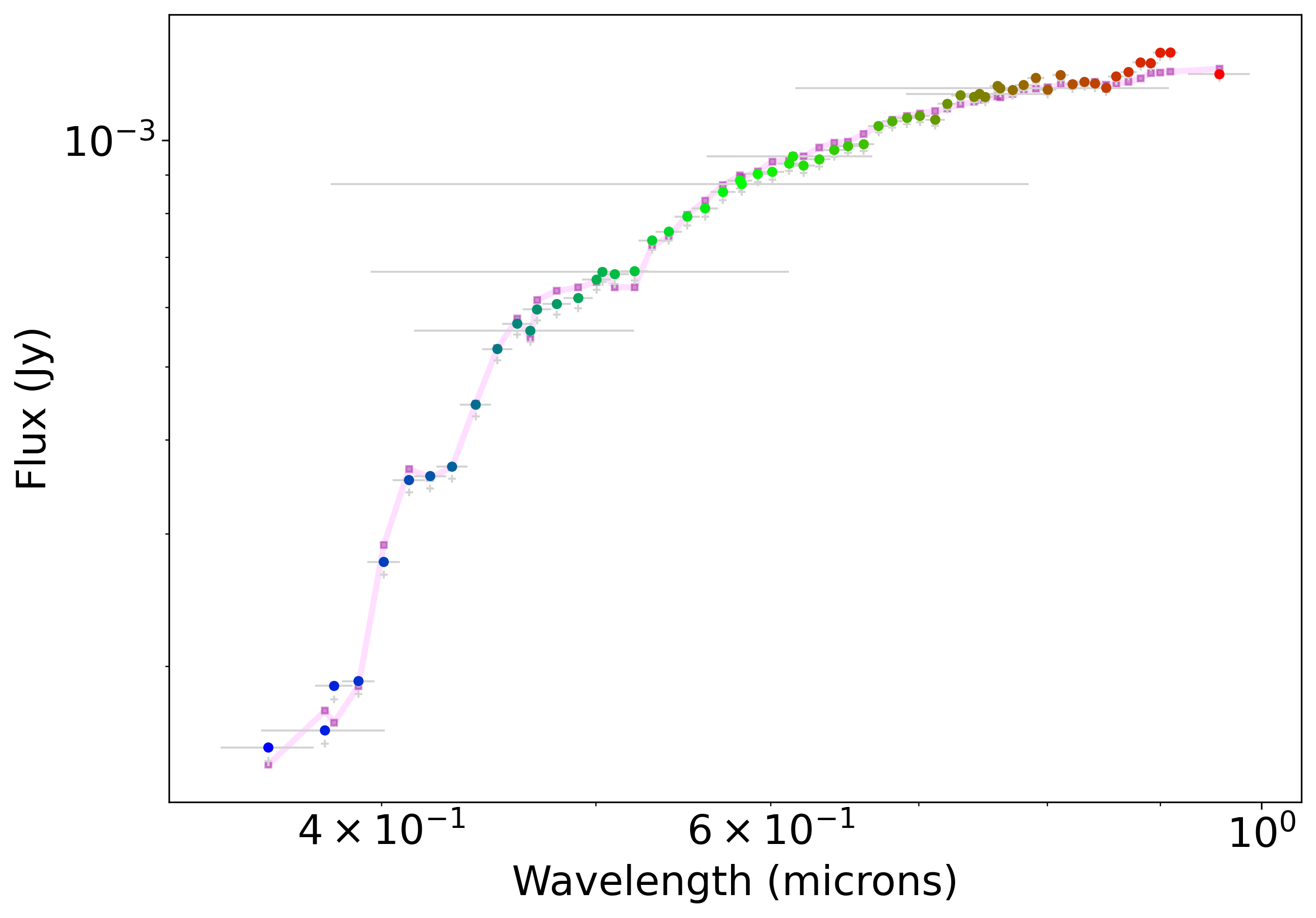}
      \caption{SEDs of \emph{Gaia} DR3 1608003977160533376 showing a standard, automated PySSED reduction (top panel, trimmed to only the optical photometry) and the improvement to the spectrophotometry from the addition of miniJPAS photometric data (bottom panel). Horizontal bars indicate the wavelength coverage of the filter, clearly separating the narrow-band and broad-band filters. Th grey plus signs indicate the observed photometry, and the round colour dots show the reddening-corrected photometry.}
      \label{fig:jpas}
\end{figure}

Figure \ref{fig:jpas} shows an example star from this dataset, indicating the improvement to spectrophotometry expected from this survey. Table \ref{tab:jpas} indicates that this star is accurately fit with or without the addition of the miniJPAS data. However, with this quality of photometry, additional science becomes possible. For example, it becomes possible to reliably disentangle temperature from both circumstellar and interstellar reddening as demonstrated in Section \ref{sec:ExtFitting} (though the very low extinction of the miniJPAS footprint prevents its sensible use here). In particular, the inclusion of data from bluer filters accurately captures the Balmer jump, allowing for greatly improved calibration of hotter stars.

\begin{table}
    \centering
    \begin{tabular}{ll}
    \hline
    Data source & $T_{\rm eff}$ (K) \\
    \hline
    LAMOST \citep{Xiang2019}    & 5498 $\pm$ 191 \\
    \emph{Gaia} DR3  & 5170$^{+29}_{-38}$ \\
    J-PAS \citep{Yang2022} & 5309 \\
    J-PAS \citep{Wang2022} & 5368 $\pm$ 7 \\
    PySSED (standard) & 5220 \\
    PySSED (with miniJPAS) & 5233 \\
    \hline
    \end{tabular}
    \caption{Table showing published effective temperatures for \emph{Gaia} DR3 1608003977160533376.}
    \label{tab:jpas}
\end{table}




\section{Conclusions}
\label{sec:Conclusions}

{\sc PySSED} represents a new public, portable code to fit stellar SEDs and extract the fundamental properties of stars. It is designed to perform this in a simple, automated way that is suitable for use by novice users and students, yet retain the power for large-scale, professional astrophysical analysis.

We have demonstrated the robustness of {\sc PySSED} in recovering stellar parameters in the cases where traditional SED-fitting codes struggle, and have highlighted circumstances that still remain problematic. The most significant of these is in regions of high extinction, where deficiencies in even the best 3D extinction cubes mean we lack an adequate measurement of interstellar extinction to remove on a per-star basis. We have demonstrated the possibility for joint fitting of extinction and temperature, while highlighting its limitations.

We have demonstrated a number of use cases of likely interest to the wider astronomical community, and shown the potential of the software as the number and variety of deep surveys increases. We have also outlined possible future applications, including uses of narrow-band spectro-photometry.

\section*{Acknowledgements}
This project has received funding from the European Union’s Horizon 2020 research and innovation programme under grant agreement No 101004214. We also acknowledge funding from UKRI/STFC through grants ST/T000414/1 and ST/X001229/1. This research has made use of the VizieR catalogue access tool, CDS, Strasbourg, France (DOI : 10.26093/cds/vizier). The original description of the VizieR service was published in 2000, A\&AS 143, 23. This research has made use of the Spanish Virtual Observatory (\url{https://svo.cab.inta-csic.es}) project funded by MCIN/AEI/10.13039/501100011033/ through grant PID2020-112949GB-I00.

\section*{Data Availability}

All astronomical data used in this work is obtained from the SVO and CDS/Vizier and is publicly available. The described version of the routine can be obtained from \url{https://github.com/iain-mcdonald/PySSED}. A version with a graphical user interface is incorporated as a data application in EXPLORE (\url{https://explore-platform.eu}), the code for which can be found at \url{https://github.com/explore-platform/s-phot}. Both versions are released under the Apache 2.0 licence. Bugs and feature requests may be reported through the standard GitHub procedures, or via e-mail with the corresponding author or the EXPLORE project, as appropriate.



\bibliographystyle{mnras}
\bibliography{pyssed} 







\bsp	
\label{lastpage}
\end{document}